\documentclass[superscriptaddress, aps,twocolumn, pra,showkeys]{revtex4-1}
        \usepackage{graphicx}
	\usepackage{subfigure}
	\usepackage{amsmath}%,amssymb} 
	\usepackage[bbgreekl]{mathbbol}
	\usepackage{makeidx}
	\usepackage{amsfonts}
	\usepackage{bm}
	\usepackage{dsfont}
	\usepackage{amssymb}
	\usepackage[ansinew]{inputenc}
	\usepackage{subfigure}
	\usepackage{float}
	\usepackage{epsfig}
	\usepackage[dvipsnames]{xcolor}
	\usepackage[colorlinks,hyperindex]{hyperref}
	\hypersetup
	{
		colorlinks,%
		citecolor=black,%
		linkcolor=black,%
		urlcolor=black,%
	}

\makeindex

\begin{document}

\title{Impurities  and other defects in correlated lattice electrons: Friedel oscillations and interference patterns }

\author{Banhi Chatterjee}
\email{banhi.chatterjee@ijs.si}
\affiliation{Jo\v{z}ef Stefan Institute, Jamova 39, SI-1000 Ljubljana,
  Slovenia}
\affiliation{Institute of Physics (FZU), Czech Academy of Sciences, Na
  Slovance 2, 182 21 Prague, Czech Republic}
\author{Jan Skolimowski}
\affiliation{Scuola Internazionale Superiore di Studi Avanzati (SISSA), Via Bonomea 265, I-34136 Trieste, Italy}
\author{Krzysztof Byczuk}
\affiliation{Institute of Theoretical Physics, Faculty of Physics, University of Warsaw, ul. Pasteura 5,  02-093 Warszawa, Poland}

\date{\today}

\begin{abstract}

  We study interference patterns and  Friedel oscillations (FO) due to  scattering  from two or more localized impurities and scattering  from extended inhomogeneities in the two-dimensional lattice systems of interacting fermions. 
  Correlations between particles are accounted for by using an approximate method based on the real-space dynamical mean-field theory and a homogeneous self-energy approximation (HSEA), where the site-dependent part of the self-energy is neglected.
  We find that the interference maxima and minima change  systematically as we vary the relative distance between the two impurities. 
  At the same time, the increase of the interaction does not shift the position of interference fringes but only reduces their intensities. 
  A comparison with the single impurity cases clearly shows  complex patterns in FO fringes induced by additional multiple scattering processes.  
  In the case of an extended step like potential the system becomes more homogeneous when the interaction  increases. 
  FO and interference patterns are not present in the Mott insulating phase in both single and  many impurity models.

  %Our theoretical study provides promising initial insights and motivate a further realistic modeling to investigate the role of interstitial defects, embedded impurities, ad-atoms on the surfaces, and surface irregularities in materials with different degrees of electronic correlation. \bcnote{},\jsnote{}, \kbnote{}

\end{abstract}


\maketitle 

\section{Introduction}
\label{intro}

The presence of an external inhomogeneous potential or localized defects in metallic systems modulates the electronic charge density around these lattice imperfections due to scatterings of the electrons near the Fermi surface. 
These charge density oscillations are known as the Friedel oscillations (FO) and are mostly visible at low temperatures \cite{Friedel52,Friedel58,alloul_friedel_2012}. 
FO occur in real materials due to the presence of ad-atoms, interstitial defects, or surface irregularities after cleavage. 
The studies of FO can be significant for a wide range of systems as we discussed in \cite{1742-6596-592-1-012059,byczuk2019t,chatterjee2019real}. 
To mention a few, FO have been observed experimentally in quantum corals, metal surfaces like Cu (111), semiconductor surfaces like GaAs-(111) around point defects using scanning tunneling microscopy (STM) at around 4-5 K \cite{Crommie93,Eigler90,Kanisawa01,Binnig82}. 
FO have been seen to produce an asymmetry in the quantum transport at the interface of mono-layer and bi-layer graphene which can be used as an application in novel quantum switching  devices \cite{clark2014energy}. 
It has been demonstrated that the FO, due to the topological defects in the carbon nanotubes, are important for understanding properties like selective dot growth, magnetic interaction through carbon nanotubes and optical spectroscopy of interface states using the tight-binding model \cite{chico}. 
Kolesnychenko \textit{et al.} observed, using a scanning tunneling microscopy (STM),  anisotropic FO while studying the surface electronic structure of transition metal Cr(001) produced by the cleavage of a single crystal having surface areas, where impurity concentrations slightly exceeded the bulk concentration due to the existence of a high dopant zone in the crystal \cite{kolesnychenko2005surface}.  \\

In order to understand  theoretically the FO in Cr or other transition metals, which belongs to the class of correlated electronic systems, it is important to consider the Coulomb interaction between the electrons. 
FO have been studied in one-dimensional (1d) interacting fermionic chain using several theories, e.g., the bosonization method and the density-matrix renormalization group \cite{egger1995friedel,schuster2004local}, or the Bethe-Ansatz local-density approximation \cite{vieira2008friedel}. 
The Fermi liquid theory for two-dimensional  (2d) and three-dimensional (3d) systems \cite{simion2005friedel} was applied to investigate FO. 
The FO induced by non-magnetic impurities in 2d Hubbard model in the presence of interactions have been studied using the slave-boson technique, which involves the static renormalization of the impurity potential \cite{ziegler1998friedel}. 
FO seen around the Kondo screening cloud in the presence of magnetic impurities using the t-matrix formalism have been reported in \cite{affleck_friedel_2008}.  \\

The dynamical mean-field theory (DMFT) is considered as an advanced and suitable technique to capture the effects of correlations particularly around the Mott metal to insulator transition which is significant for describing the compounds with partially filled d and f shells, e.g., transition metals and their oxides \cite{vollhardt2012Dynamical,vollhardt_investigation_1993,kotliar2004strongly,byczuk2008dynamical,titvinidze_dynamical_2012,0953-8984-9-35-010}.   
A real space extension of DMFT (R-DMFT) is needed to treat strongly correlated inhomogeneous systems \cite{Helmes08,Snook08,potthoff1999metallic,suarez2020two}. \\

%%%%%%%%%%%%%%%%%%%%%%%%%%%%%%%%%%%%%%%%%%%%%%%%%%%%%%%%%%%%%%%%%%%%%%%%%%%%%%%%%%%%%%%%%%%%%%%%%%%%%%%%%%%%%%%%%%%

 In our previous work we investigated the behavior of FO in models of correlated  lattice systems in the metallic and Mott insulating phases in the presence of a single impurity potential using the R-DMFT \cite{1742-6596-592-1-012059,chatterjee2019real}. 
 We reported that the oscillations get damped with increasing the interaction, and disappear at the Mott transition and beyond it in the Mott insulating phase.  \\

In reality, a single-site impurity potential could be a too idealized model when we deal with inhomogeneities on surfaces of true materials. 
There are usually more than one defect or contamination. 
One even encounters extended inhomogeneities and interface effects in multi-layered nano-structures, for example as discussed  in Ref.~\cite{freericks2006transport}. 
Grosu \textit{et al.} studied the problem of FO in a 1d noninteracting electron gas in the presence of two impurities, modeled by a double delta function separated by a finite distance, using a linear response theory \cite{grosu2008friedel}. 
They showed that the presence of the second impurity significantly changes the density oscillations (changing the positions of the maxima and minima) depending on the distance between the impurities. The studies of two impurity scattering have been further extended to a 1d interacting fermionic system using the bosonization method in \cite{liu2003two}. 
The scattering and quasiparticle interference from two and multiple magnetic impurities adsorbed on 2d and 3d interacting hosts have been probed using the t-matrix formalism and the numerical renormalization group \cite{derry2015quasiparticle,mitchell2015multiple}. However, in these studies the interference effects are discussed in the local density of states in the presence of interactions and not in the particle density oscillations.  \\

We thus see that the studies of FO in the presence of two or multiple impurities have been conducted mostly for  1d interacting systems, while an attempt to understand real materials requires models in two and higher dimensions.
 Moreover, the behavior of FO in the Mott insulating regime for models with many imperfections has not been addressed. 
 Also the current state of knowledge lacks any quantitative treatment of the screening and interference effects in systems with interparticle interactions.\\
 
 A proper description of FO in real materials with strong electronic correlations demands a  realistic modeling combining the density functional theory within the local density approximation and  DMFT (LDA+DMFT) \cite{schuler2016many,kolorenvc2015electronic} and its extension in real space. Such techniques are computationally non-trivial in the presence of inhomogenities, particularly if it is not just an ad-atom but an impurity atom embedded in the host, e.g. Cr atom in a Pb surface \cite{choi2017mapping}. The translational invariance of the lattice is broken in such a case.\\  
 
Motivated by this state of art we extend our earlier study \cite{chatterjee2019real} by adding to the simple one-band Hubbard model  various types of impurity potentials and inhomogeneities.  The correlations are treated by using an approximate self-energy based on DMFT where the site-dependent part of the self-energy is neglected. We name this approximation the homogeneous self-energy approximation (HSEA) and discuss it in  Sec.~II.
We investigate both non-interacting and interacting two-dimensional finite lattice systems.\\ 

In this paper we address the following questions: 
(a) How do the FO change due to the interference effects when we introduce the second impurity? 
(b) How does the interference change when we vary the distance between the impurities and switch on the interaction? 
(c) How does the whole picture change if we generalize the two impurities to multiple impurities scattered over the lattice or the extended inhomogeneity? 
(d) Do we see any interference effect or FO for any of these models of impurity potential in the Mott insulating phase?\\

Our investigation shows that the interaction reduces the interference pattern in FO and the impurity screening. 
 However, the interaction does not alter the position of the interference maxima and minima for the particle-hole symmetric case.\\

The paper is organized as follows.
 in Sec.~II we discuss  lattice models and methods used to solve them. 
We introduce there physical quantities describing physical properties of these  systems. 
Afterwards in Sec.~III we present our numerical results for: (a) two impurities, (b) multiple impurities, (c) a chain of impurities or a domain wall, and (d) extended inhomogenity.  
Finally, in Sec.~IV we conclude our results and provide an outlook.\\
           
     %%%%%%%%%%%%%%%%%%%%%%%%%%%%%%%%%%%%%%%%%%%%%%%%%%%%%%%%%%%%%%%%%%%%%%%%%%%%%%%%%%%%%%%%%%%%%%%%%%%%%%%%%%%%%%%%%%%%%%%%%%%%%%%%%%%%%%%     

\section{Models and formalism}      

\subsection{Models} 

    We consider the one-band Hubbard model \cite{hubbard_electron_1963, gutzwiller1963effect, kanamori1963electron} with an external inhomogeneous potential  
	\begin{equation}
	 H = \sum_{ij, \sigma} t_{ij}\  \hat{a}_{i\sigma}^\dag\ \hat{a}_{j\sigma} +\sum_{i\sigma} V_{i\sigma}\ \hat{a}_{i\sigma}^\dag\ \hat{a}_{i\sigma} + U \sum_{i} \hat{n}_{i\downarrow}\hat{n}_{i\uparrow}  , 
	\label{hubbard}
	\end{equation}
    where $\hat{a}_{i\sigma}$ ($\hat{a}_{i\sigma}^{\dag}$) is the annihilation (creation) fermionic operator with spin $\sigma$ on the $i^{th}$ lattice site, $t_{ij}$ is the hopping matrix element between the $i^{th}$ and $j^{th}$ sites with $t_{ii}=0$. The second term describes the external (inhomogeneous) potential given by $V_{i\sigma}$ which is assumed to be real. The third term models the local part of the electronic interaction between two fermions with opposite spins located on the same lattice site.\\
    
    We consider a two dimensional square lattice with  the  number of lattice sites $N_{L}=31^2$ (the size of the lattice is $31\times31$) and the following models of the external (inhomogeneous) spin independent potential:
    \begin{itemize}
    
     \item Two single site impurities placed either along the diagonal direction of the lattice or  along a vertical direction of the lattice for different relative distances. Mathematically, $V_{i}=V_{01}\delta_{i i_{01}}+V_{02}\delta_{i i_{02}}$. 
    
     \item A more general case where several impurities are irregularly distributed over  various lattice sites. This aims to model a contaminated surface with various dopant zones or interstitial defects. Mathematically, $V_{i}=V_{01}\delta_{i i_{01}}+V_{02}\delta_{i i_{02}}+V_{03}\delta_{i i_{03}}+..$.

      \item A chain of impurities placed along the diagonal and vertical directions of the lattice aimed to model a domain wall or an interface. In freshly cleaved samples such domain walls can be found connecting large lattice inhomogeneties and can be observed experimentally using the STM.

     \item A step like potential or extended inhomogeneity across the lattice aimed to describe inhomogneous surface irregularities after a cleavage. Mathematically,  $V_{i\sigma}= V_{0}\Theta (X_i-X_0)$, where $X_0$ is the horizontal lattice coordinate where the steplike potential begins (i.e., the Heaviside function $\Theta(x)$ is non-zero).

    \end{itemize}
    
     %%%%%%%%%%%%%%%%%%%%%%%%%%%%%%%%%%%%%%%%%%%%%%%%%%%%%%%%%%%%%%%%%%%%%%%%%%%%%%%%%%%%%%%%%%%%%%%%%%%%%%%%%%%%%%%%%%%%%%%%%%%%%%%%%%%%%%%

    \subsection{Method}
    
\begin{figure*}
\centering
\begin{minipage}{0.4\textwidth}
\centering
\includegraphics[width=0.8\textwidth]{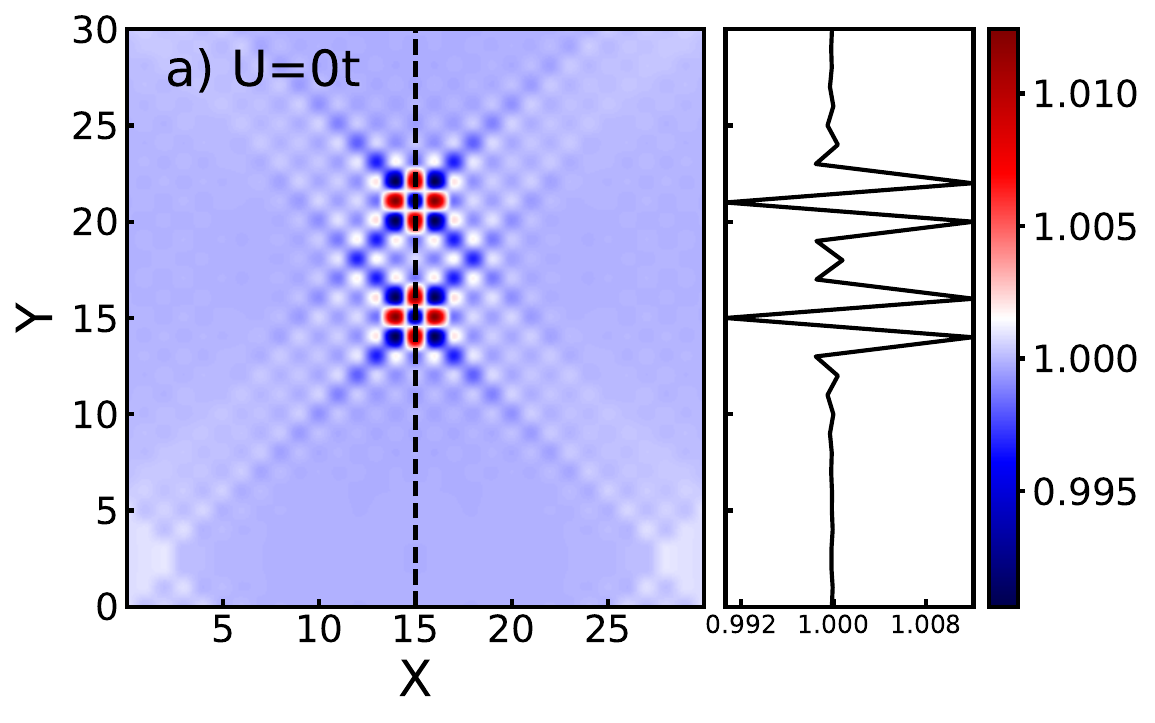} % first figure itself
%        \caption{first figure}
\end{minipage}
\begin{minipage}{0.4\textwidth}
\centering
\includegraphics[width=0.8\textwidth]{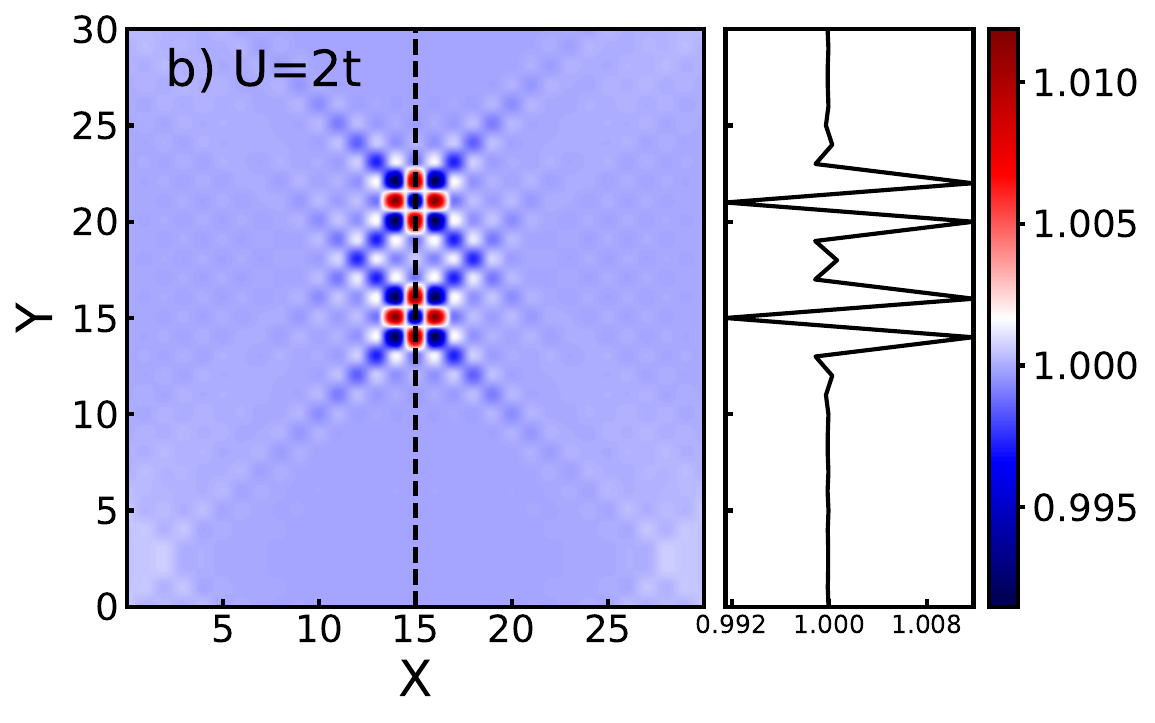} % first figure itself
%        \caption{first figure}
\end{minipage}\hfill
\begin{minipage}{0.4\textwidth}
\centering
\includegraphics[width=0.8\textwidth]{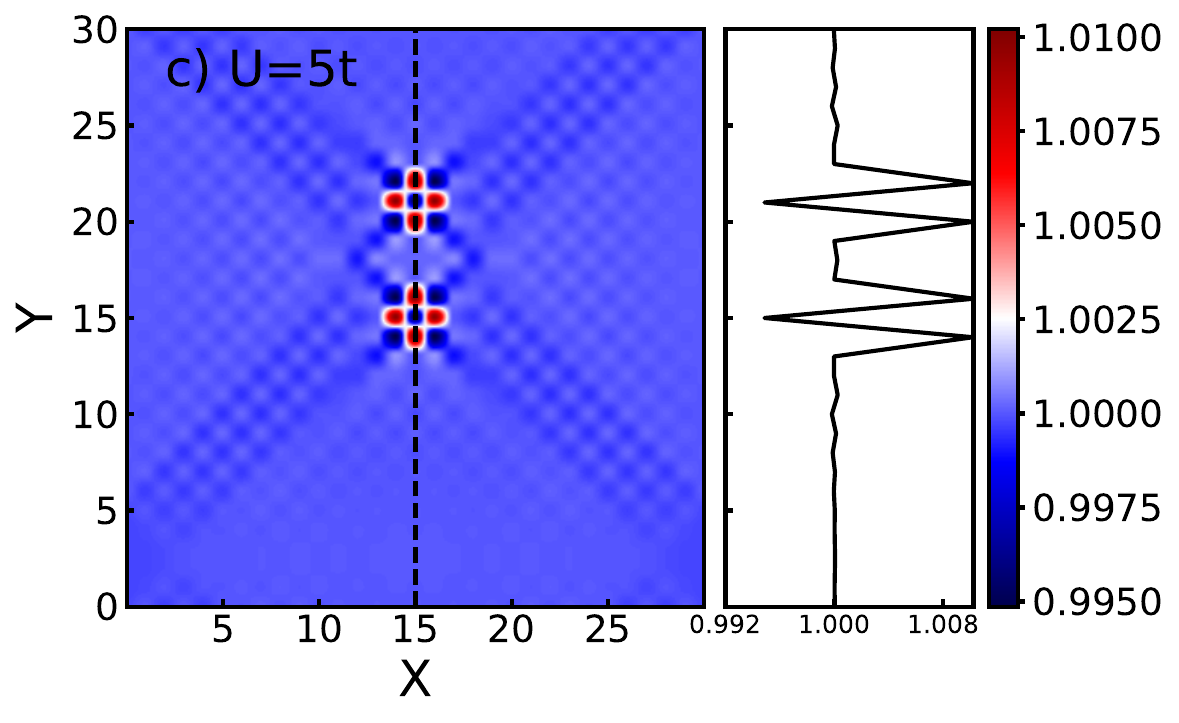} % second figure itself
%        \caption{second figure}
\end{minipage}
\begin{minipage}{0.4\textwidth}
\centering
\includegraphics[width=0.8\textwidth]{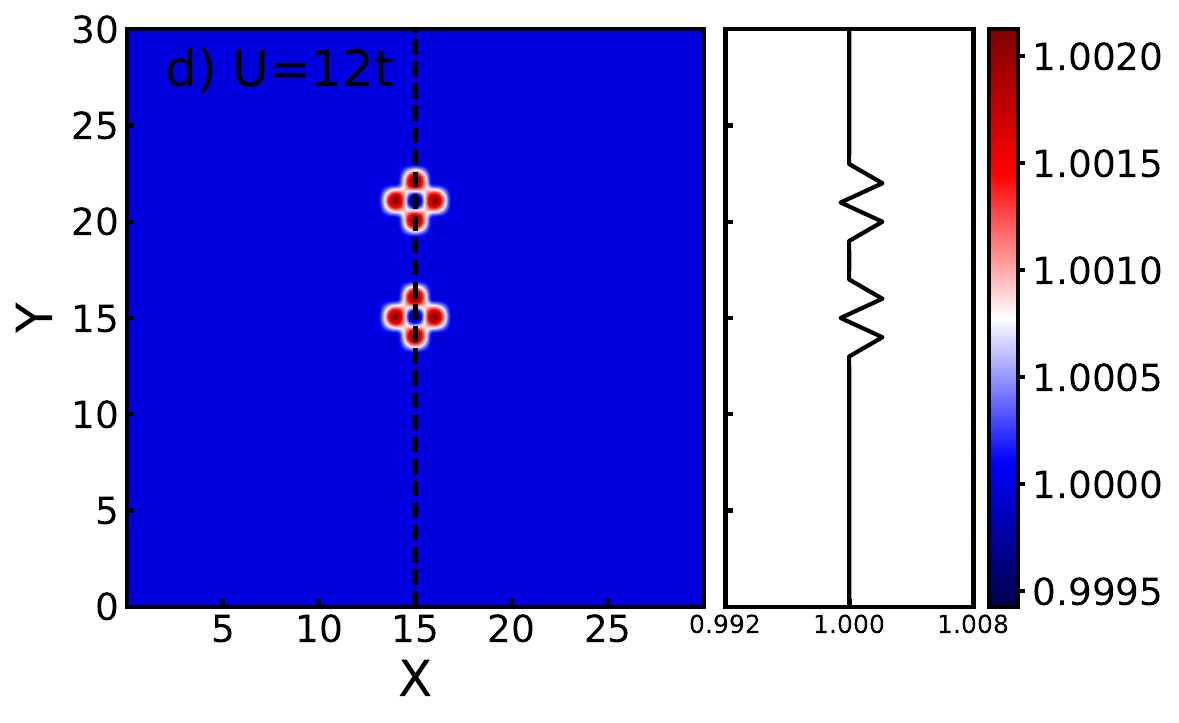} % second figure itself
%        \caption{second figure}
\end{minipage}
\caption{Friedel oscillations (FO) in particle densities ($\bar{n}_{i}$) due to two impurities of equal magnitude, i.e. $V_{01}=V_{02}=24t$,  placed at $\vec{R}_{01}=(15a,15a)$ and $\vec{R}_{02}=(15a,22a)$, respectively, along the vertical axis of the ($31\times31$) square lattice. We show the change in interference pattern of FO due to scattering from the two impurities for a) $U=0t$, b) $U=2t$, c) $U=5t$, and
d) $U=12t$. The insets (on the right) show oscillations along the vertical line passing through the impurities. The color scale is spanned in between the highest and the lowest values of the density in the system. The color scale changes for different $U$ since the minimal value of the density increases with it. We do not show the actual densities on the impurity sites, which are substituted by the second lowest values in the plots. Otherwise the impurity contribution would overshadow any contributions in $\bar{n}_{i}$.} 
\label{2dtwoimpvert}
\end{figure*}

\begin{figure} [ht!]
\centering
\includegraphics[width=0.3\textwidth]{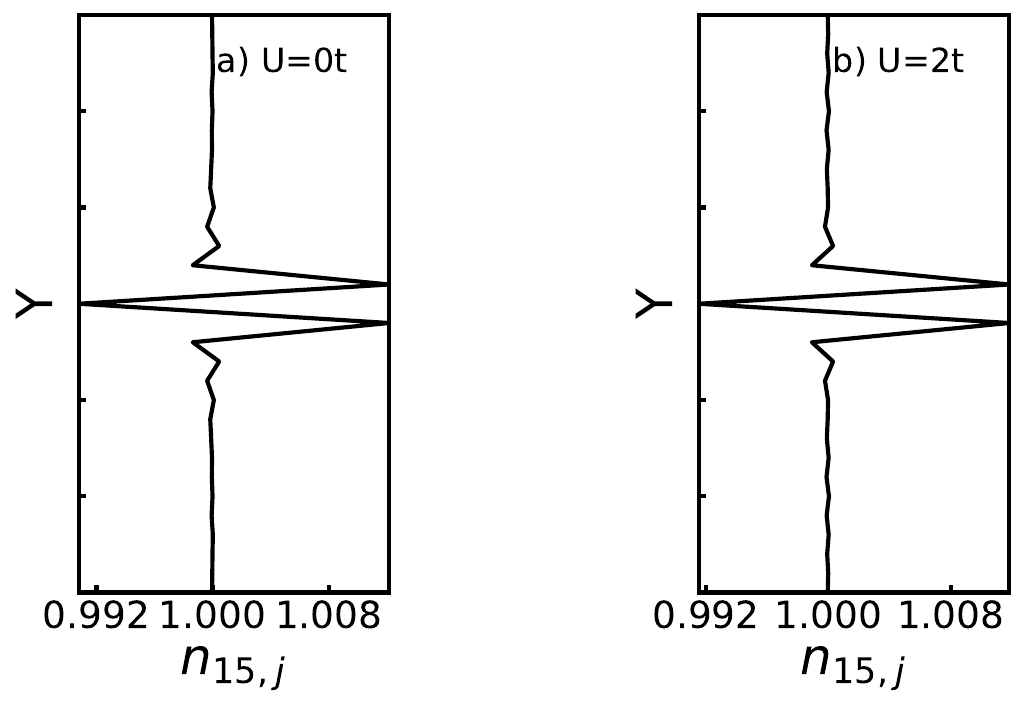} 
\qquad
\includegraphics[width=0.3\textwidth]{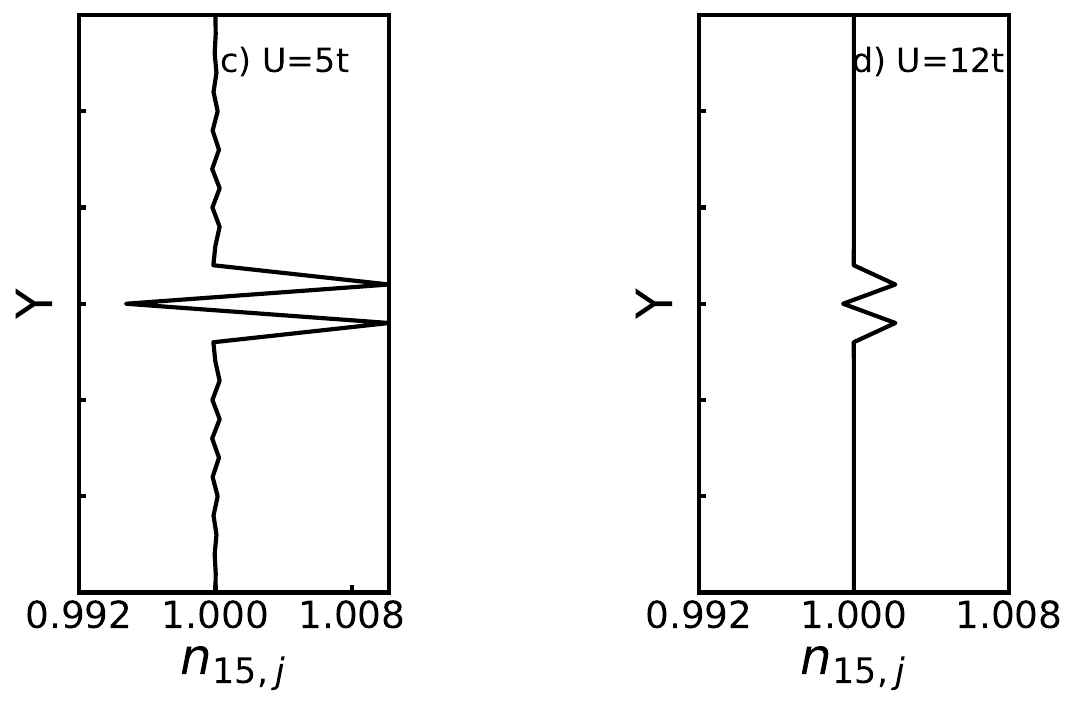}  
\caption{FO in particle denisties ($\bar{n}_{i}$)  in the square lattice in the presence of a single impurity $V_{0}=24t$ in the centre ($\vec{R}_{0}=(15a, 15a)$). All other model parameters and plotting style is the same as in Fig.~\ref{2dtwoimpvert}. We show a) $U=0t$, b) $U= 2t$, c) $U=5t$, and d) $U=12t$. We only show the vertical line passing through site containing the impurity for a comparison with Fig~\ref{2dtwoimpvert}. A complete density profile for the single impurity case is available in \cite{chatterjee2019real}.}
\label{denmapsimp}
\end{figure}

     All single particle properties of the system are obtained from the retarded Green's function obtained by inverting the matrix Dyson equation \cite{rickayzen1980green} in the lattice position space    
   \begin{equation}
    \mathbf{G}(z)= [(z+\mu)\mathbf{1}-\mathbf{t}-\mathbf{V}-\Sigma(z)\mathbf{1}]^{-1},
    \label{greenfunction}
   \end{equation}
where $z= \omega+ i0^+$ is the energy approaching  the real axis from above and  $\mu$ is the chemical potential ($\hbar=1$). 
$\Sigma(z)$ is the site independent homogeneous part of the self-energy which approximates the effect of correlations and is calculated using the DMFT for the same parameters of the corresponding homogeneous Hubbard Hamiltonian.
 Hereafter, all matrices are expressed in bold faced notation. 
 The non-diagonal matrix $\bf t$ corresponds to the hopping amplitudes $t_{ij}$ and the diagonal matrix ${\bf V}$ reflects the on-site inhomogeneous potential $V_i$.   The unity matrix is written as $\bf 1$.  \\
   
   In the DMFT the self-energy is diagonal in lattice site indices and accounts for all local dynamic correlation effects. In case of homogeneous lattice systems, all lattice sites are equivalent. In this case, the self-energy is computed  by mapping a lattice site into an effective single impurity Anderson model (SIAM) and solving it by using standard techniques like a continuous time quantum Monte Carlo, an exact diagonalization, a numerical renormalization group method (NRG),  etc. However, in the presence of external impurities, translational invariance of the lattice is broken and the lattice sites are non-equivalent. Hence,  it is essential now to solve the SIAM separately at each lattice site and the local self-energy becomes explicitly site dependent. In other words, in an inhomogeneous system the self-energy has a homogeneous part due to the electron-electron interactions and an inhomogeneous part due to the contribution from the interaction and the external impurities. Owing to the site dependent part of the self-energy, the result of the impurity potential in the system is not static but effectively dynamic \cite{byczuk2019t}. Ideally, in order to get the full picture of a correlated inhomogeneous system we should consider both the homogeneous and inhomogeneous part of the self energy solving the full real-space DMFT (R- DMFT) equations self-consistently\cite{byczuk2019t}.\\
   
 Unfortunately, deciphering  the full R-DMFT equations is computationally exhaustive, especially for higher-dimensional systems with a large number of lattice sites. Hence, as a first approximation  we omit the inclusion of inhomogeneous, site-dependent part of the self-energy in our present studies to obtain some initial insights on the behavior of the system with correlations. We call this approximation homogeneous self-energy approximation  \cite{chatterjee2019real}. The homogeneous part of the self-energy is computed by solving the DMFT self-consistency equations for infinite homogeneous system at zero temperature and half-filling  (particle-hole symmetric case) by using NRG  method \cite{bulla_zero_1999}. This essentially implies setting the parameter $V_{i\sigma}=0$ in Eq.~\ref{hubbard}. The open-source code ``NRG Ljubljana'' \cite{vzitkonrg,  costi1990new} is used for this purpose. The computed self-energy (the HSEA) is then transferred to the real space Dyson equation (\ref{greenfunction}) containing the impurity potential $V_{i\sigma}\neq 0$ and used to obtain the one-particle Green's function. We note here that although the self-energies are computed for a homogeneous system, the Green's function is still obtained by inverting the Dyson equation containing the impurity potential in the real space and thus inhomogeneity of the system is taken into account. \\
   
   A detailed discussion of the R-DMFT and HSEA is presented in \cite{chatterjee2019real} wherein we also show that the results obtained from these two methods qualitatively agree for a single impurity potential. It might be an interesting future project to compare the results obtained from the full R-DMFT and from the HSEA for our models of the multiple  impurity potentials. 
 However, we do not expect significant changes because even in the metallic regime the FO decay quickly is space with a power law. We also observe, that a possible change in the density of particles due to a change of the self-energy by inhomogeneous potential is a higher order effect beyond the linear response regime.

%   Nevertheless, in this work we discuss the changes one may expect over HSEA on including the full R-DMFT. \bcnote{I recommend readers the previous PRB for the detailed mathematical formalism. I also address the objection of the review that two methods agree 'well' by removing the word 'well'. KB, JS> Do we need to provide more technicalities in this part?}  \\

%%%%%%%%%%%%%%%%%%%%%%%%%%%%%%%%%%%%%%%%%%%%%%%%%%%%%%%%%%%%%%%%%%%%%%%%%%%%%%%%%%%%%%%%%%%%%%%%%%%%%%%%%%%%%%%%%%%%%%%%%%%%%%%%%%%%%%%
   \subsection{Physical quantities}
   
   Once we know the Green's function of the system from Eq.~(\ref{hubbard}) and (\ref{greenfunction}) we obtain the local spectral function as 
    \begin{equation}
	A_{i \sigma}(\omega)=-\frac{1}{\pi} \text{Im} \ G_{ii\sigma}(\omega).
	\label{spectral}
	\end{equation}
    Having (\ref{spectral}) we compute the spin resolved local density of particles at zero temperature as 
 \begin{equation} \label{local_occupationhom}
   \bar{n}_{i\sigma}= \int_{-\infty}^{E_F} A_{i\sigma}(\omega)f(\omega) \,d\omega .
   \end{equation}
We consider spin rotational invariant systems, i.e., $\bar{n}_{i\uparrow}=\bar{n}_{i\downarrow}$, and the total number of particles per site is given by
	\begin{equation}
	\bar{n}=\frac{1}{N_{L}}\sum_{i=1}^{N_L} \bar{n}_{i},
	\end{equation}
    where $\bar{n}_{i}=n_{i\uparrow}+n_{i\downarrow}$.
Eq.~(\ref{local_occupationhom})  is the most relevant for our studies of FO. \\
    
  We further quantify the screening and interference effects in the system by the screening charge defined by
	  \begin{equation}
	  Z=\sum\limits_{i\sigma}\left(\bar{n}_{i\sigma}-\frac{\bar{n}_{\rm hom}}{2}\right),
	 % \left[\frac{\mathbf{p_{i}^{2}}}{2m}+U_{per}(\mathbf {r})+U_{ext}(\mathbf {r})\right]
	  \label{screen1}
	\end{equation}
    where the summation runs over all the lattice sites and $\bar{n}_{\rm hom}$ corresponds to the density of particles of the reference  homogeneous system (i.e., with $V_{i}=0$).

\section{Numerical Results}
\label{results}

 %   In this Section we present our numerical results.\\

    We  choose the chemical potential $\mu= U/2$ such that the homogeneous system is at half-filling, i.e. $\bar{n}=1$. In all cases the hopping amplitude $t_{ij}=t$ is only between nearest neighbors. We set $t=1$ to define the energy units and set  the lattice constant $a=1$ to define the length units in our numerical calculations. The band-width W of the system is given by $W=2zt$, where z is the co-ordination number. The system is subjected to the periodic boundary conditions with a finite number of the lattice sites $N_{L}$. We perform our simulations at  the zero temperature ($T=0$).\\
    
    The strength of electronic correlation is controlled by tuning the parameter $U$. 
    We study the 2d homogeneous system for different $U$ values and see that the Mott transition occurs at $U_{c}\approx11.5t$. 
    Hence, we choose $U=0t$,  $2t$, $5t$, and $12t$  values to represent a non-interacting lattice gas, a weakly interacting metallic phase, an intermediate interacting metallic phase, and a  Mott insulating phase of the inhomogeneous system,  respectively. \\

      \subsection {Two impurities}

\begin{figure*}
\centering
\begin{minipage}{0.4\textwidth}
\includegraphics[width=0.8\textwidth]{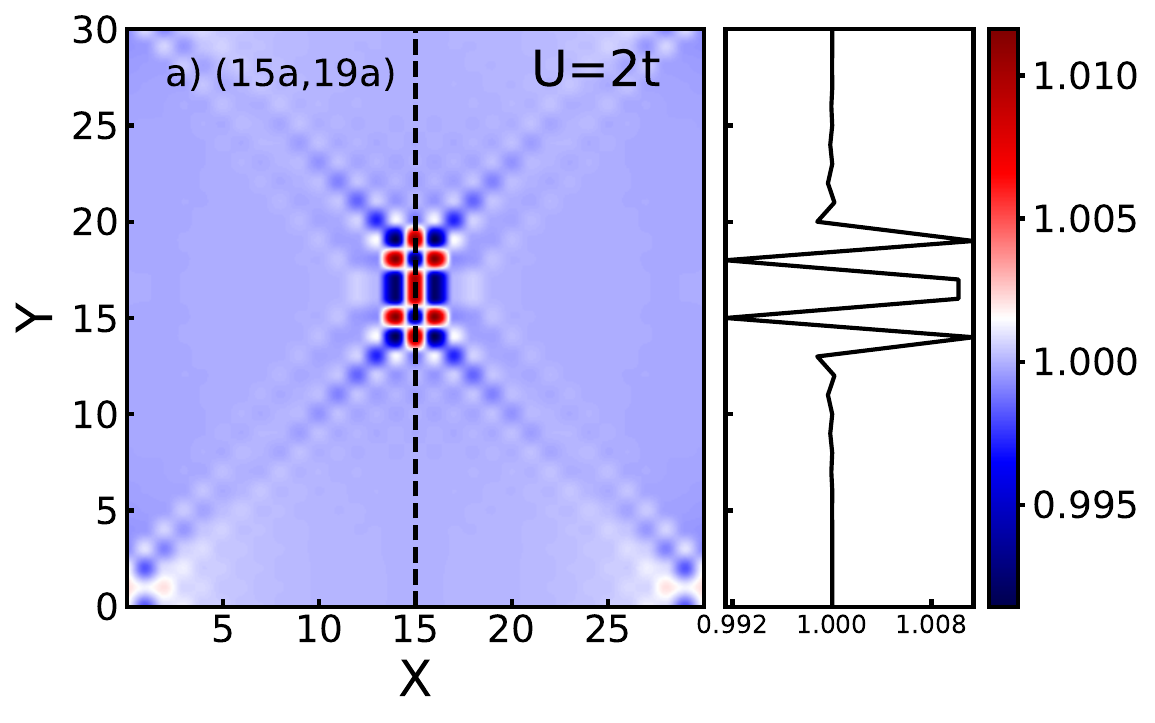} % first figure itself
%       \caption{(15a,19a)}
\end{minipage}
\begin{minipage}{0.4\textwidth}
\includegraphics[width=0.8\textwidth]{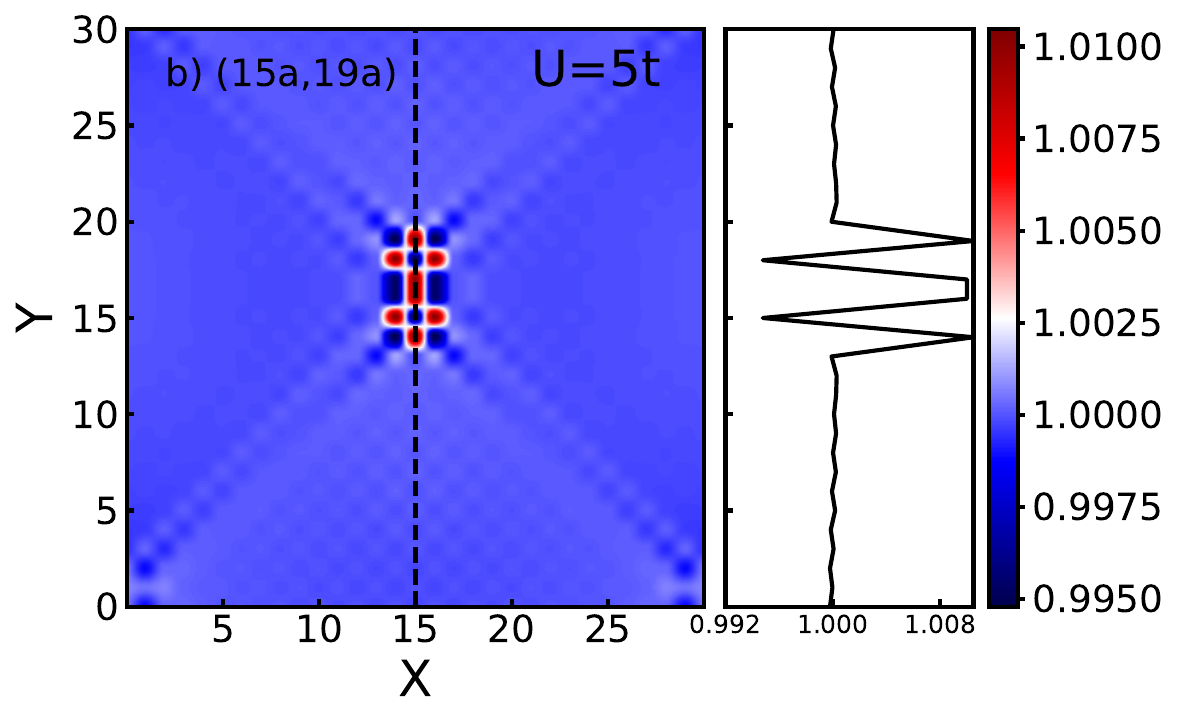} % second figure itself
%        \caption{second figure}
\end{minipage}
\begin{minipage}{0.4\textwidth}
\includegraphics[width=0.8\textwidth]{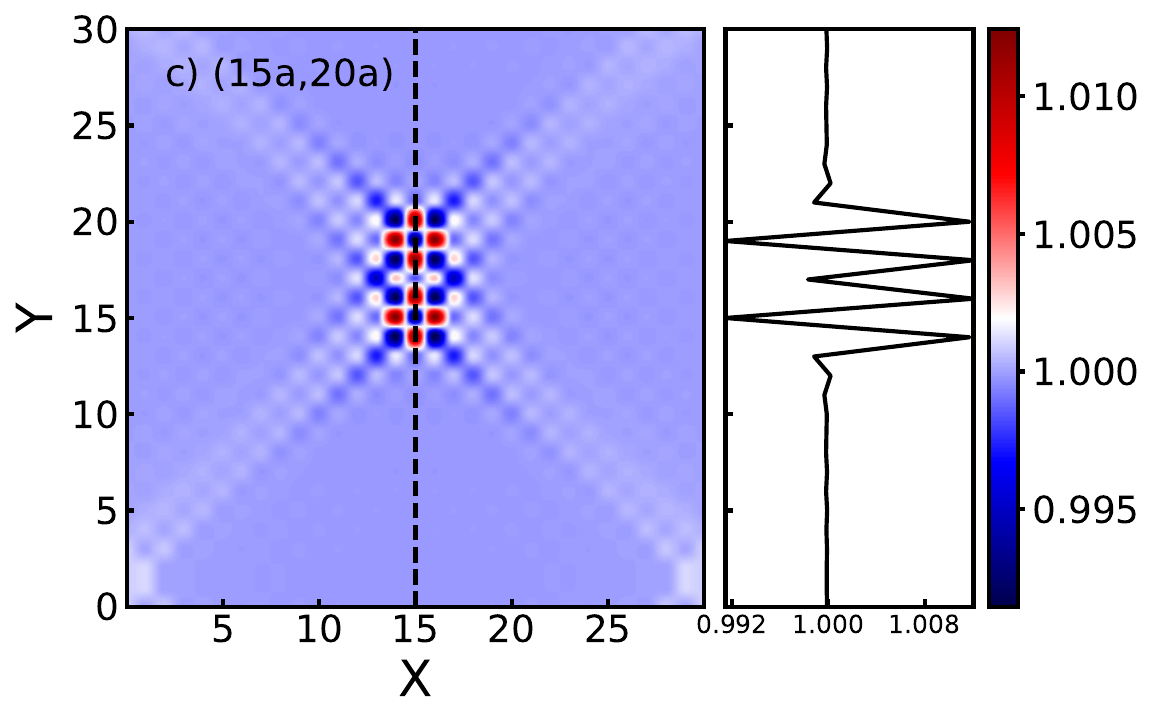} % second figure itself
%        \caption{second figure}
\end{minipage}
\begin{minipage}{0.4\textwidth}
\includegraphics[width=0.8\textwidth]{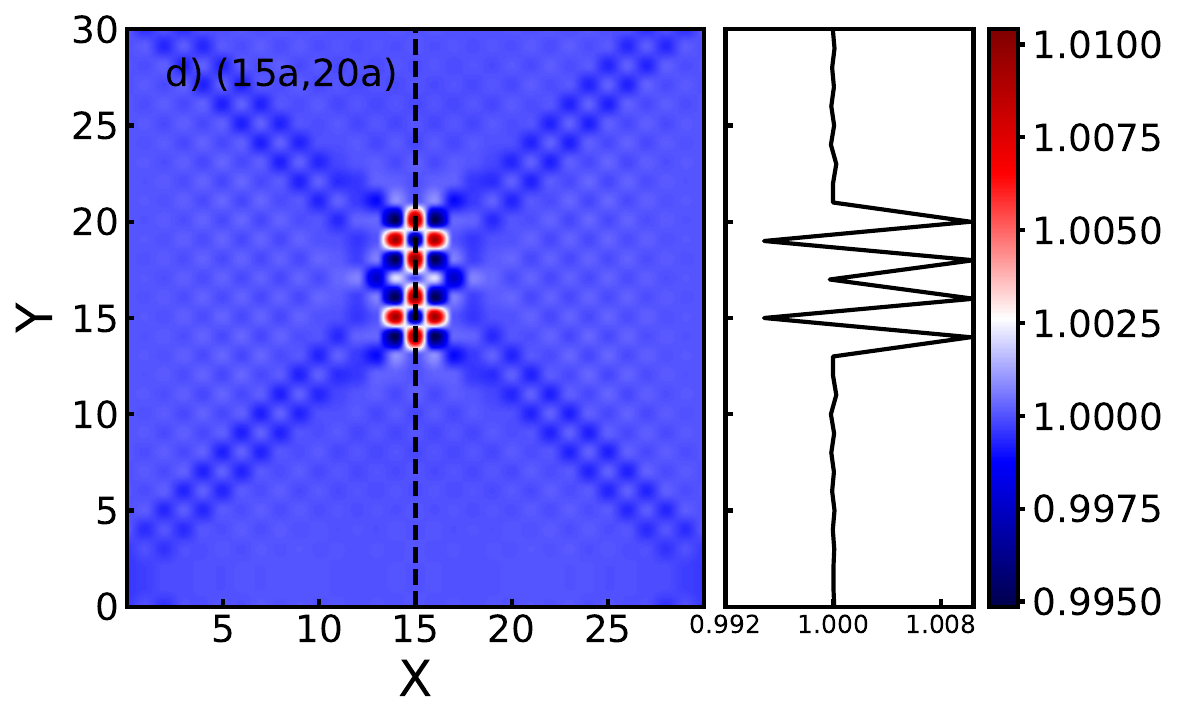} % second figure itself
%        \caption{second figure}
\end{minipage}
\begin{minipage}{0.4\textwidth}
\includegraphics[width=0.8\textwidth]{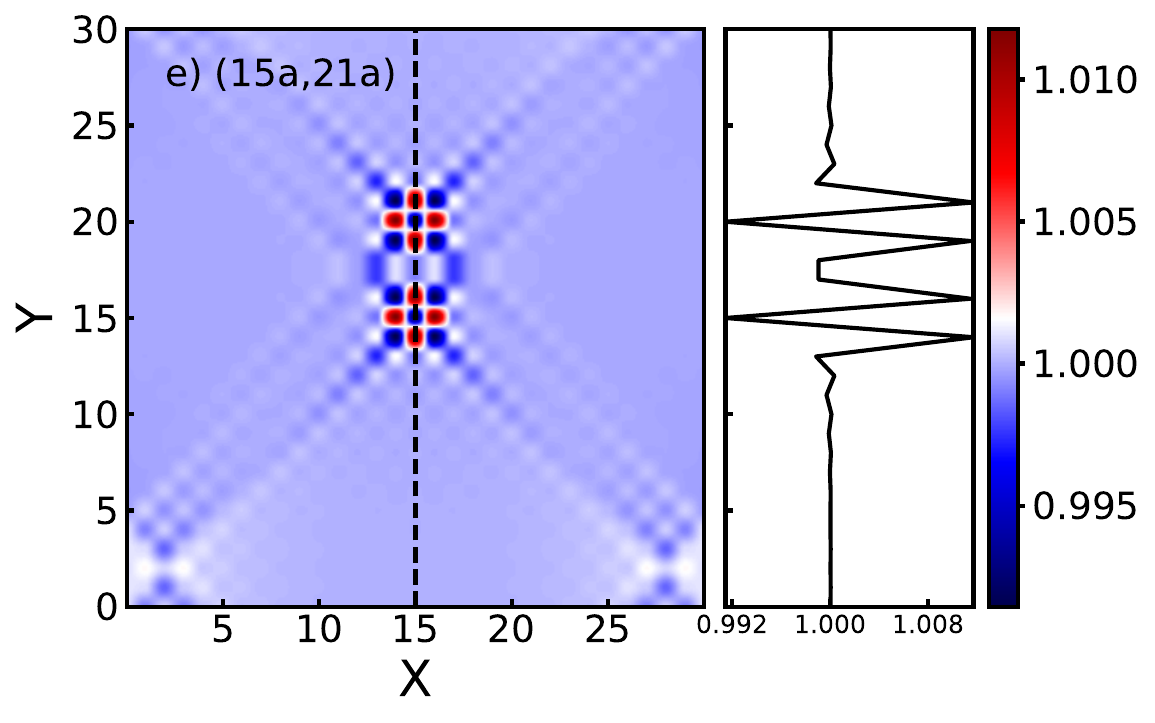} % second figure itself
%        \caption{second figure}ma 
\end{minipage}
\begin{minipage}{0.4\textwidth}
\includegraphics[width=0.8\textwidth]{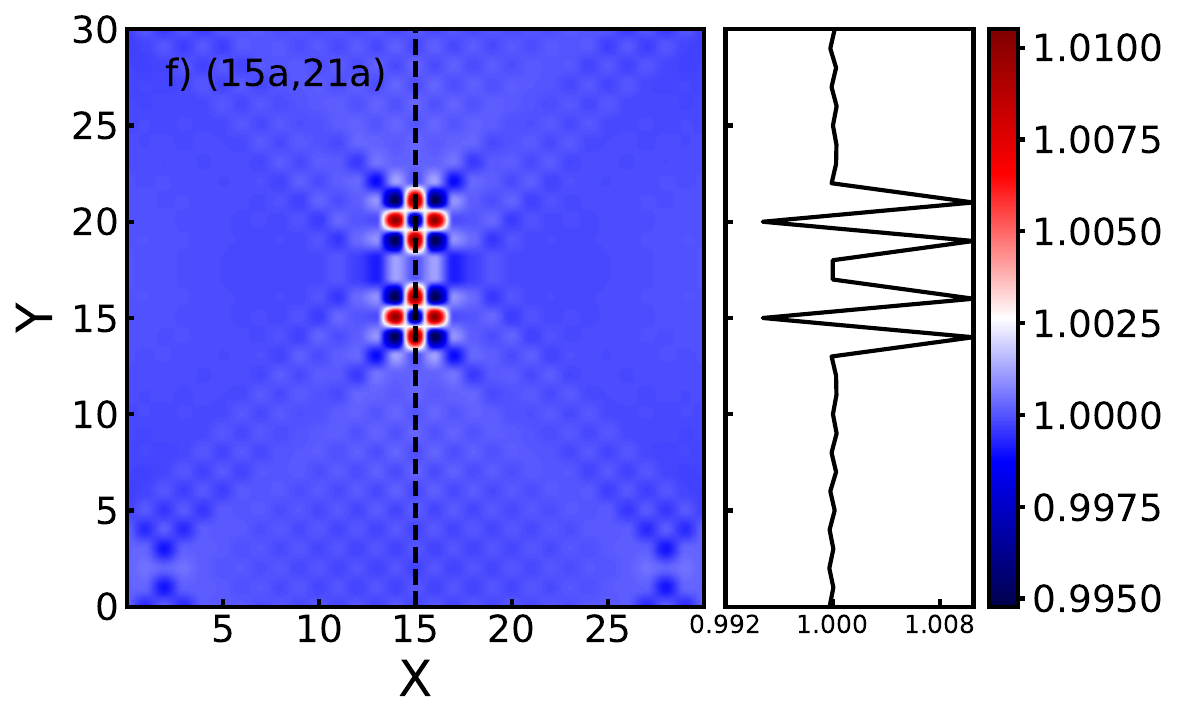} % second figure itself
%        \caption{second figure}
\end{minipage}
\begin{minipage}{0.4\textwidth}
\includegraphics[width=0.8\textwidth]{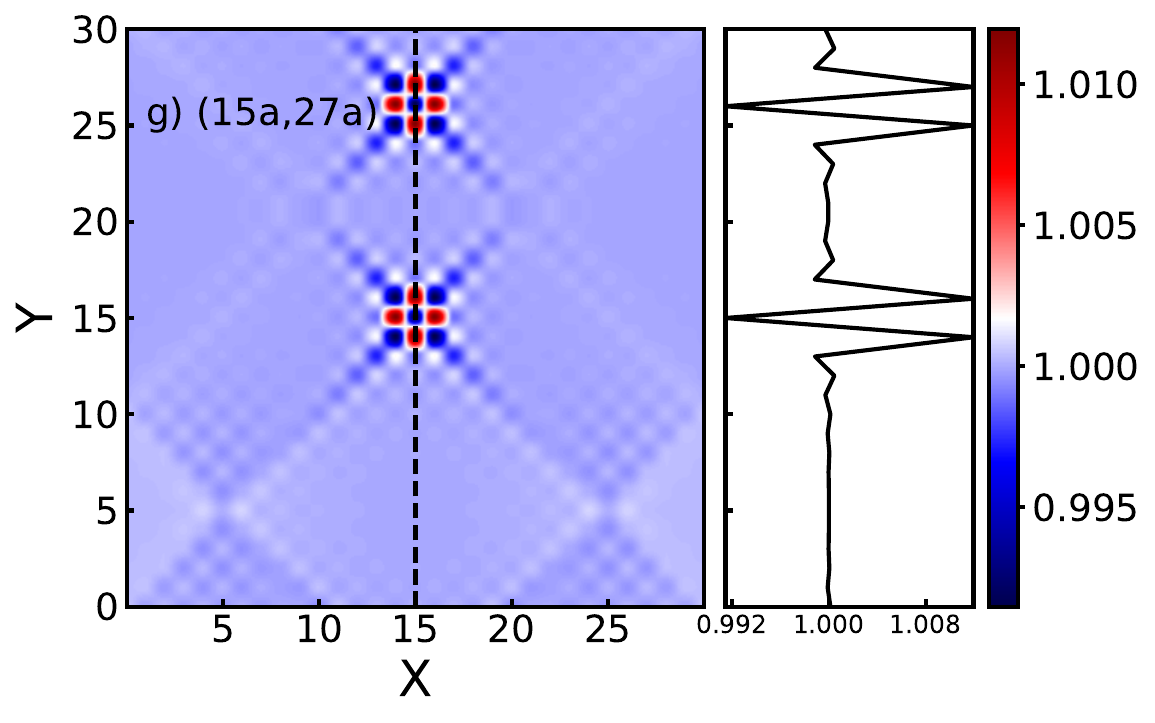} % second figure itself
%        \caption{second figure}
\end{minipage}
\begin{minipage}{0.4\textwidth}
\includegraphics[width=0.8\textwidth]{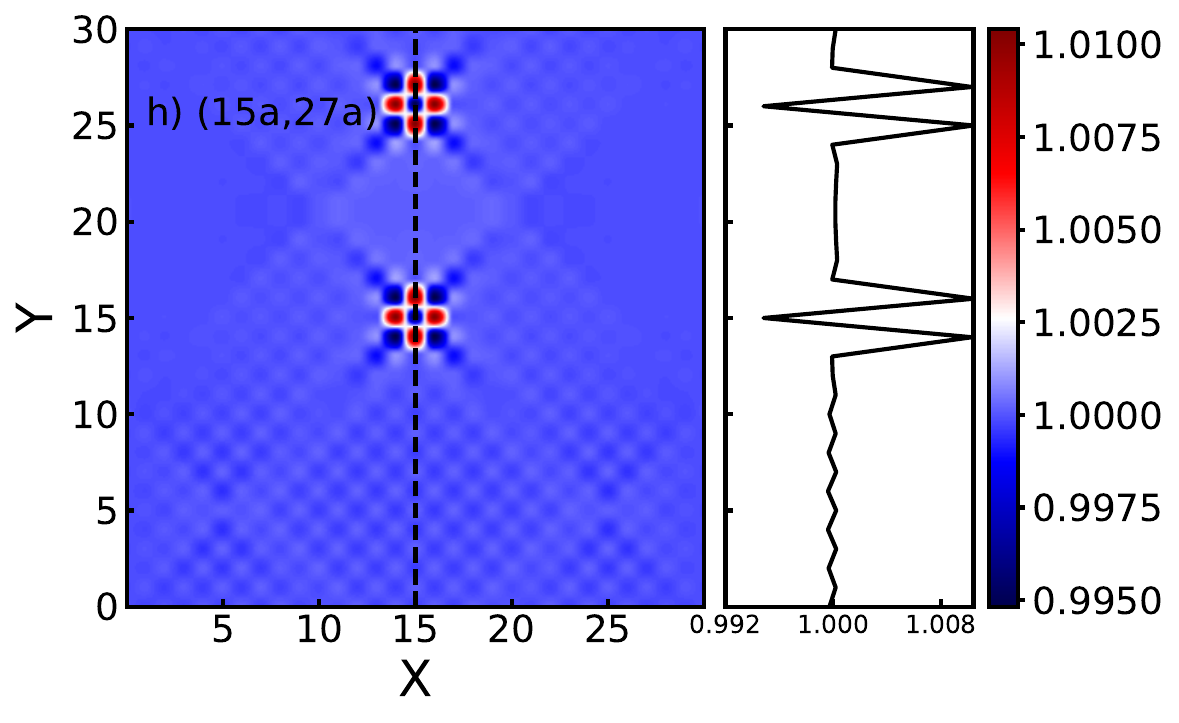} % second figure itself
%        \caption{second figure}
\end{minipage}
\caption{The variation in the interference patterns in the FO of particle densties ($\bar{n}_{i}$) for different relative distances between the two  impurities of equal magnitude $V_{01}=V_{02}=24t$. The first impurity  $V_{01}$ is kept fixed at $\vec{R}_{01}=(15a, 15a)$ and the position of the second impurity is shifted along the vertical line and placed at $\vec{R}_{02}=(15a, 19a)$, $\vec{R}_{02}=(15a, 20a)$, $\vec{R}_{02}=(15a, 21a)$, $\vec{R}_{02}=(15a,27a)$ (from top to bottom, respectively). We show the interaction $U=2t$ (a,c,e,g), and $U=5t$ (b,d,f,h). All other model parameters, and the plotting style are the same as in Fig.~\ref{2dtwoimpvert}.} 
\label{2dtwoimpvertdist}
\end{figure*}

     We start with the  case where two impurities are present in the system. 
In Fig.~\ref{2dtwoimpvert} we show the interference patterns  in FO due to two impurities of equal magnitude $V_{01}=V_{02}=24t$ placed in the lattice sites $(15a, 15a)$ and $(15a, 22a)$ along the vertical direction of  the square lattice for the non-interacting system and the interacting system with  $U=2t$, $ 5t$, and $12t$. 
We further compare  with the case where only a single impurity is present in the system in Fig.~\ref{denmapsimp}. 
On comparing Fig.~\ref{2dtwoimpvert} and Fig.~\ref{denmapsimp} the interference effect induced by the second impurity is visible. 
We see that within the HSEA the interaction does not change the position of the interference maxima and minima but reduces their heights and intensities as seen in $U=2t$ and $5t$ cases. 
This is analogous to the damping of FO with the interactions in the single impurity case. 
Such behavior is attributed to the particle-hole symmetry in the system. 
The natures of the non-interacting system and the weakly-interacting one with $U=2t$ are very similar. 
No interference effects and FO are visible in the Mott insulating phase  at $U=12t$. 
The disappearence of the FO in the Mott insulating phase  is due to  a gap opening around the Fermi level, and the Fermi edge cutoff, responsible for oscillatory behavior, is absent. \\

     In order to investigate how the interference maximum and minimum changes as we vary the relative distance between the impurities, in Fig.~\ref{2dtwoimpvertdist} we fix the position of the first impurity at $(15a, 15a)$ and vary the position of the second impurity to: $(15a,19a)$, $(15a,20a)$, $(15a, 21a)$, and $(15a,27a)$. 
     We show only the cases for $U=2t$, and $ 5t$ since the behavior of the non-interacting system is very similar to the $U=2t$ one and no interference effects were seen in the system with  $U=12t$. 
     We see the occurrence  of a minimum for $(15a,19a)$, and a maximum for $(15a, 20a)$. 
     Beyond a certain cross-over distance, e.g., for $(15a,27a)$ the interference effects are negligible and the inhomogeneities behave independently as in the diluted impurity regime. 
     Again, on comparing the cases $U=2t$ and $U=5t$ we see that the interaction does not change the position of the maximum and minimum but reduces their height. 
     In other words, we conclude that the interaction reduces the interference effects.\\
     
\begin{figure} [ht!]
\centering
\includegraphics[width=0.25\textwidth]{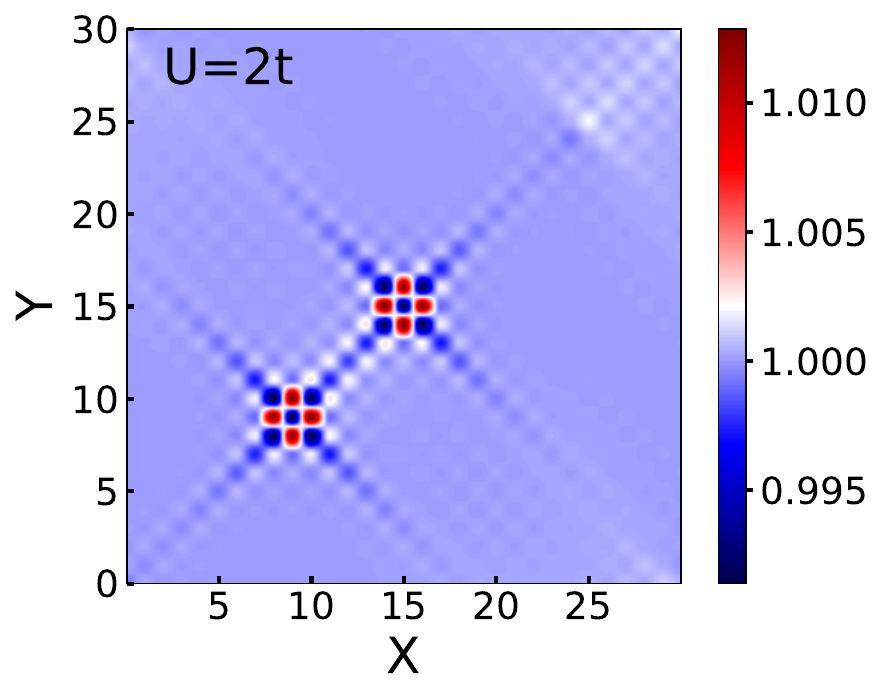} 
\qquad
\includegraphics[width=0.25\textwidth]{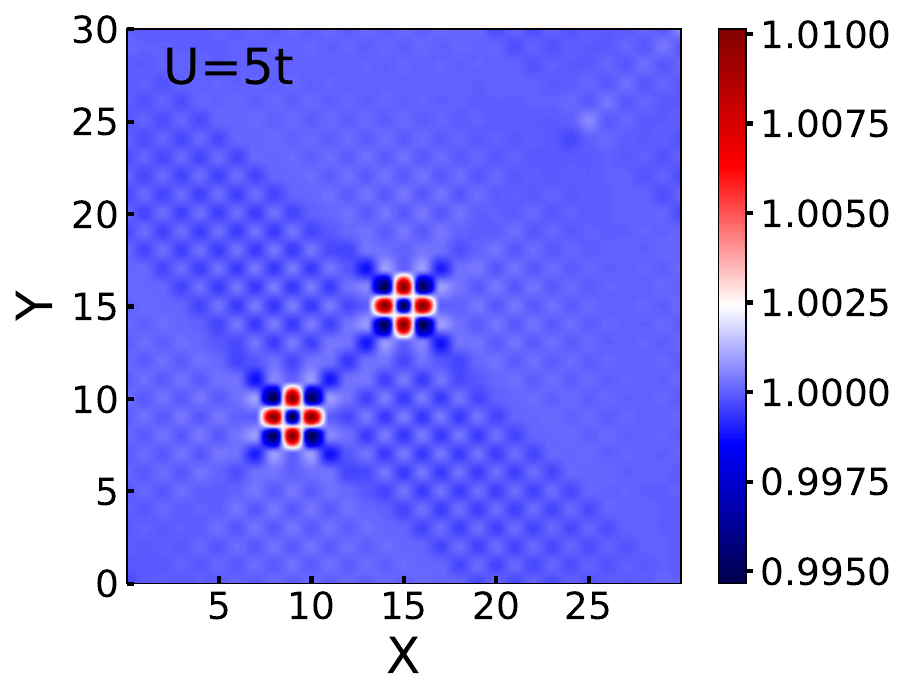} 
\caption{Friedel oscillations (FO) in particle densities ($\bar{n}_{i}$) in the presence of two impurities of equal magnitude, i.e., $V_{01}=V_{02}=24t$,  placed at $\vec{R}_{01}=(9a,9a)$ and $\vec{R}_{02}=(15a,15a)$, respectively, along the diagonal direction of the square lattice. The interactions $U=2t$ (upper panel) and $U=5t$ (lower panel). The model, all other parameters, and the plotting style are the same as in Fig.~\ref{2dtwoimpvert}.}
\label{2dtwoimpdiagsame}
\end{figure}

     We now place two impurities of equal magnitude $V_{01}=V_{02}=24t$  along the diagonal direction of the lattice at sites $(10a, 10a)$ and $(15a, 15a)$ and show the cases for $U=2t$, and $U=5t$ in Fig.~\ref{2dtwoimpdiagsame}. Comparing Fig.~\ref{2dtwoimpdiagsame} and Fig.~\ref{2dtwoimpvertdist} we already see that the interference patterns are qualitatively different when the impurities are placed along the diagonal. Particularly, for  $U=5t$ alternate regions of high and low density along the diagonal of the lattice are clearly visible. The interference  pattern in the interstitial region  between the two impurities is the most dominant. No FO or scattering interference effects are visible for $U=12t$.\\
     
\begin{figure} [ht!]
\centering
\includegraphics[width=0.4\textwidth]{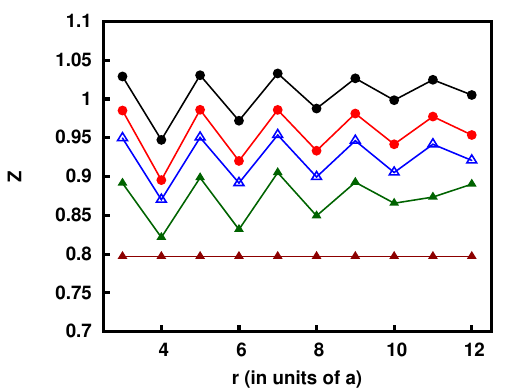} 
\qquad
\includegraphics[width=0.4\textwidth]{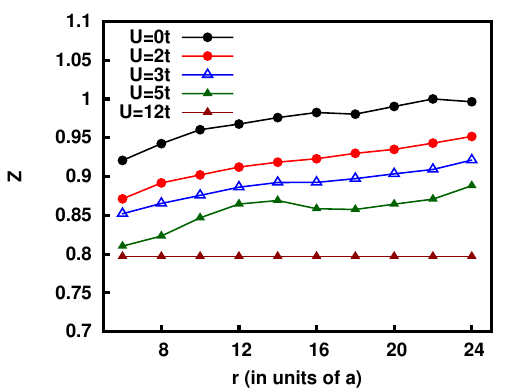}  
\caption{Variation of the screening charge defined by Eq.~(\ref{screen1}) with relative (Manhattan) distance  between the impurities ($V_{01}= V_{02}= 24t$). The impurities are placed in the vertical (diagonal) directions in the upper (lower) panel. Different values of the parameter U corresponds to the different interactions. We show $U=0t$, $2t$, $5t$, and $12t$. The legends for the plot in the upper panel (not shown) is same as that of the lower panel. }
\label{2dtwoimpscreen}
\end{figure}

In order to get a quantitative description of the interference effects with the relative distance between the impurities we study the dependance of the screening charge as defined by Eq.~(\ref{screen1}) for different positions of the second impurity, i.e.,  when placed along the vertical or along diagonal directions as shown in Fig.~\ref{2dtwoimpscreen} (top and bottom panels) respectively. 
When the two impurities are placed along the vertical line, an oscillatory behavior is seen in the screening charge, i.e.,  maxima (minima) appear when the impurities are separated by an odd (even) number of lattice sites both for the non-interacting and interacting systems. 
%It shows the alternate constructive and desctructive interference for the even/odd spacing. 
The screening charge does not change with distance and reaches its constant residual value in the Mott phase ($U=12t$) again confirming the absence of any FO or interference effects in this regime.\\

The oscillatory behavior in screening charge with the distance is absent when the impurities are placed along the diagonal direction as seen in Fig. \ref{2dtwoimpscreen} (bottom panel). 
Along the diagonal direction the Manhattan distance between the two sites is always even and, hence, there are always interference minima in the FO. 
If one compares the evolution of Z for even lattice spacings in the upper panel with the same ones in the lower panel they almost match perfectly. The screening charge reaches the same constant residual value for $U=12t$ like in the case when the impurities are placed along the vertical direction.
In both  cases, at any given distance the screening charge reduces with the increasing interactions, which is in agreement with the case when only a single impurity is present in the system \cite{chatterjee2019real}. \\

%%%%%%%%%%%%%%%%%%%%%%%%%%%%%%%%%%%%%%%%%%%%%%%%%%%%%%%%%%%%%%%%%%%%%%%%%%%%%%%%%%%%%%%%%%%%%%%
%%%%%%%%%%%%%%%%%%%%%%%%%%%%%%%%%%%%%%%%%%%%%%%%%%%%%%%%%%%%%%%%%%%%%%%%%%%%%%%%%%%%%%%%%%%%%%%         

       \subsection {Multiple impurities}
       
%%%%%%%%%%%%%%%%%%%%%%%%%%%%%%%%%%%%%%%%%%%%%%%%%%%%%%%%%%%%%%%%%%%%%%%%%%%%%%%%%%%%%%%%%%%%%%%%%%%%%%%%%%%%%      
% \begin{figure} [ht!]
% \centering
% \includegraphics[width=0.2\textwidth]{fig11a.pdf} 
% \qquad
% \includegraphics[width=0.2\textwidth]{fig11b.pdf} 
% \qquad
% \includegraphics[width=0.2\textwidth]{fig11c.pdf}  
% \qquad
% \includegraphics[width=0.2\textwidth]{fig11d.pdf}
% \caption{FO in particle denisties ($\bar{n}_{i}$)  in the presence of five impurities each of magnitude $V_{1}=V_{2}=V_{3}=V_{4}=V_{5}=10t$ 
% randomly scattered over the lattice  sites located at:  $\vec{R}= (3a, 3a)$, $(20a, 5a)$, $(5a, 20a)$, 
% $(25a, 22a)$, and $(17a, 28a)$ of the square lattice. We show the interference in FO due to the impurities for a) a non-interacting system (U=0t) and the system with the interactions b) $U=2t$, c) $U=5t$, and d) $U=12t$. All other model parameters and the plotting style are the same as in Fig.~\ref{2dtwoimpvert}.
% }
% \label{2dmultimp}
% \end{figure}
%%%%%%%%%%%%%%%%%%%%%%%%%%%%%%%%%%%%%%%%%%%%%%%%%%%%%%%%%%%%%%%%%%%%%%%%%%%%%%%%%%%%%%%%%%%%%%%%%%%%%%%%%%%%%%%
           
    \begin{figure} [ht!]
\centering
\includegraphics[width=0.4\textwidth]{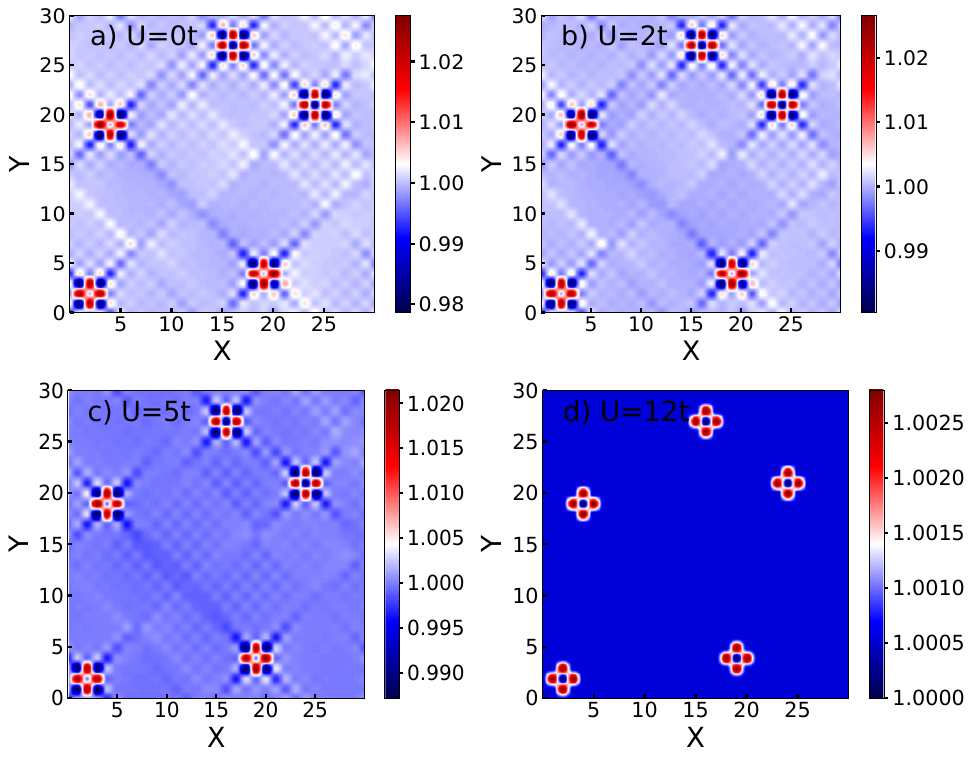} 
\caption{FO in particle denisties ($\bar{n}_{i}$)  in the presence of five impurities each of magnitude $V_{1}=V_{2}=V_{3}=V_{4}=V_{5}=10t$ distributed on the lattice  sites located at:  $\vec{R}= (3a, 3a)$, $(20a, 5a)$, $(5a, 20a)$, 
 $(25a, 22a)$, and $(17a, 28a)$ of the square lattice. We show the interference in FO due to the impurities for a) a non-interacting system (U=0t) and the system with the interactions b) $U=2t$, c) $U=5t$, and d) $U=12t$. All other model parameters and the plotting style are the same as in Fig.~\ref{2dtwoimpvert}}
\label{2dmultimp}
\end{figure}  

        We now move to a more complex case where we extend our studies to several impurities randomly distributed over the surface. This aims to model a contaminated surface of a transition metal in the presence of dopant/defects, e.g. the Cr 001 surface in \cite{kolesnychenko2005surface}. We use the square lattice, also to predict the behavior on the surfaces of 3d systems,  exploiting the fact that lower dimensions can also mimic the higher dimensions in the DMFT approximation due to the momentum independence in the self-energy. This feature has also been exploited in \cite{chatterjee2019real}.\\

In Fig. \ref{2dmultimp}, we consider five impurities each of magnitude $V_{0}=10t$,  distributed on the lattice at sites: 
$(3a, 3a)$, $(20a, 5a)$, $(5a, 20a)$, $(25a, 22a)$, $(17a, 28a)$ for the non-interacting system and systems with  $U=2t$, $U=5t$, and $U=12t$. 
We see oscillations around the impurities together with a complex interference pattern (like a checkerboard) in the interstitial spaces between the impurities for the non-interacting system, and systems with $U=2t$, $5t$. 
Interference effects get localized around the impurities with the increasing interactions (cf., $U=5t$). 
No FO are observed in the Mott insulator for ($U=12t$). 
This is in agreement to our previous studies where a single impurity or two impurities are present in the system. 
Thus we conclude that, at least within HSEA the absence of FO and any interference effects due to scattering in the Mott insulating phase is rather universal irrespective of the model of the inhomogeneous potential.\\
        
    %%%%%%%%%%%%%%%%%%%%%%%%%%%%%%%%%%%%%%%%%%%%%%%%%%%%%%%%%%%%%%%%%%%%%%%%%%%%%%%%%%%%%%%%%%%%%%%%%%%%%%%%%%%%%%%%%%%%%%%%%%%%%%%%%%%%%%%%%%%%%%%%%%%%%%%%%%%%%%%%%%%%%%%%%%%%%%%%%%%%%%%%%%%%%%%%%%%%%%%%%%%%%%%%%%%%%%%%%%%%%%

        \subsection {A chain of impurities}

\begin{figure} [ht!]
 \centering
 \includegraphics[width=0.3\textwidth]{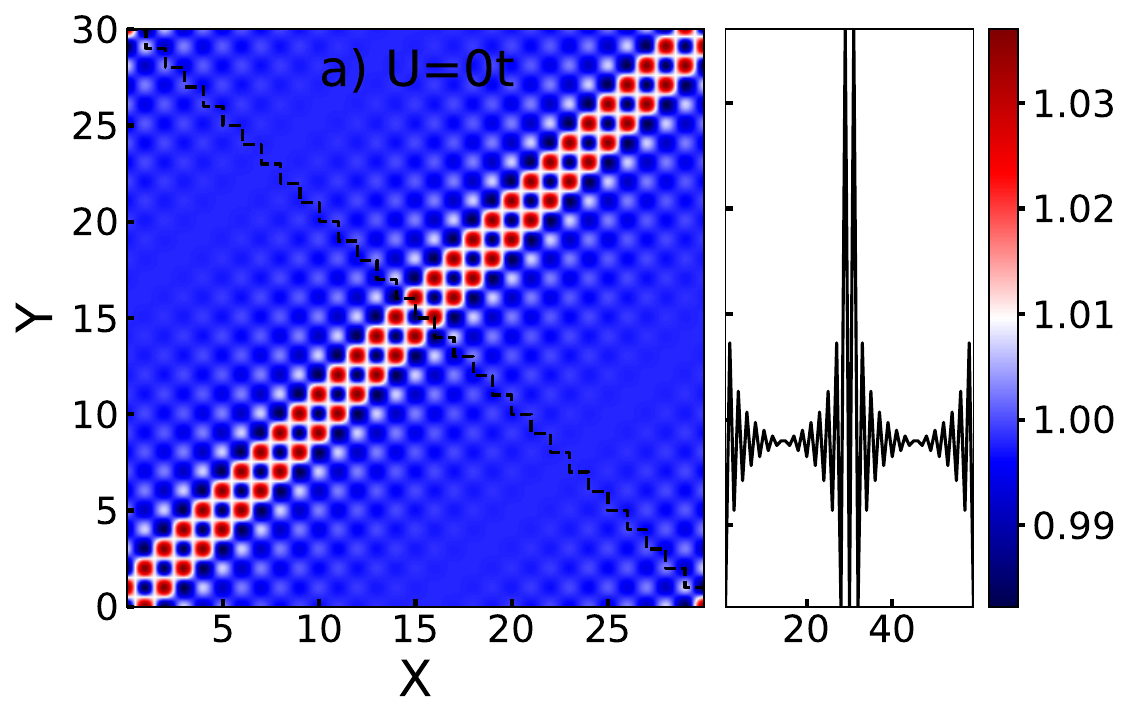} 
 \qquad
 \includegraphics[width=0.3\textwidth]{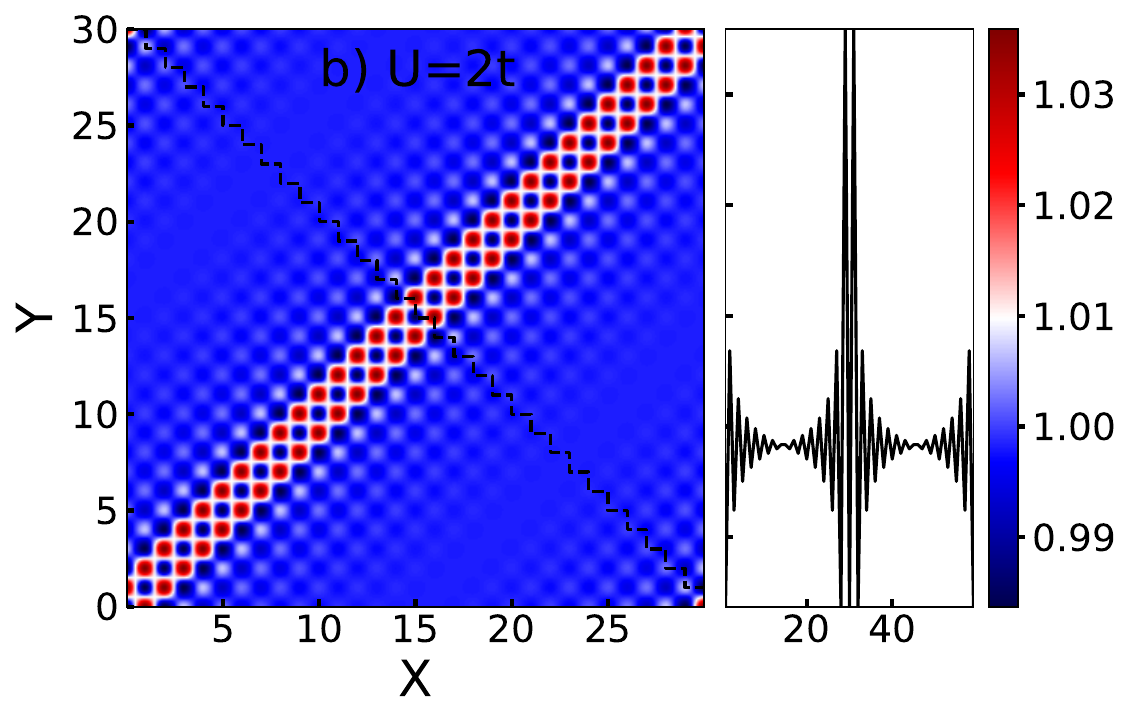}  
 \qquad
 \includegraphics[width=0.3\textwidth]{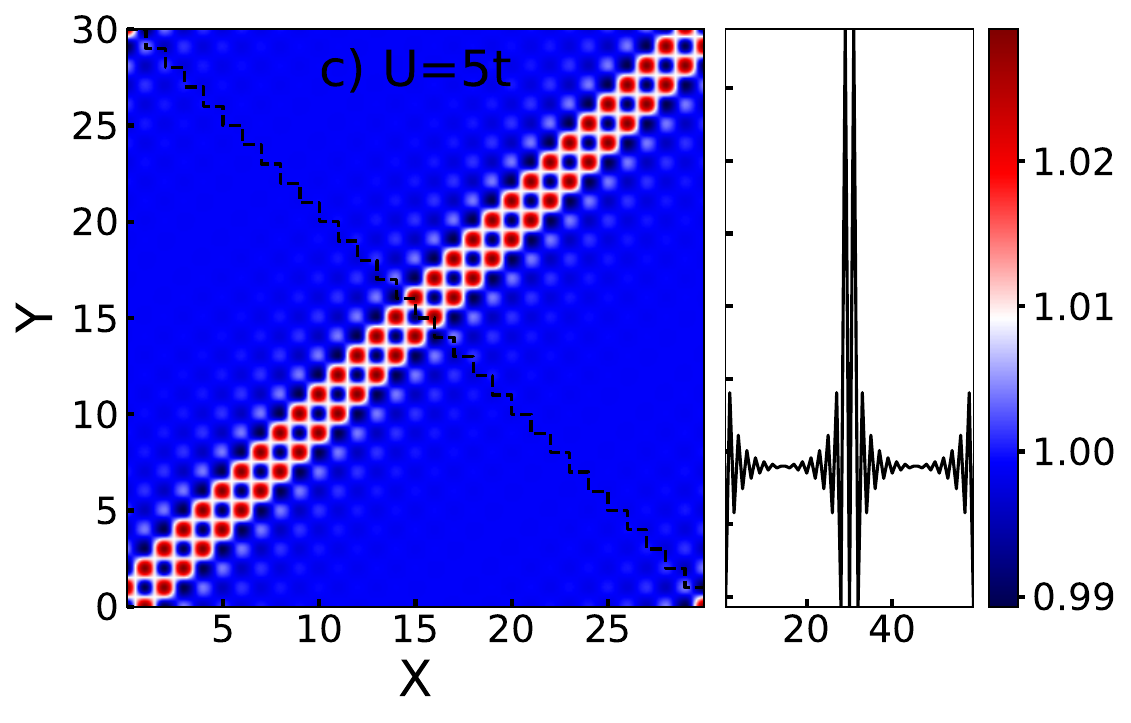} 
 \qquad
 \includegraphics[width=0.3\textwidth]{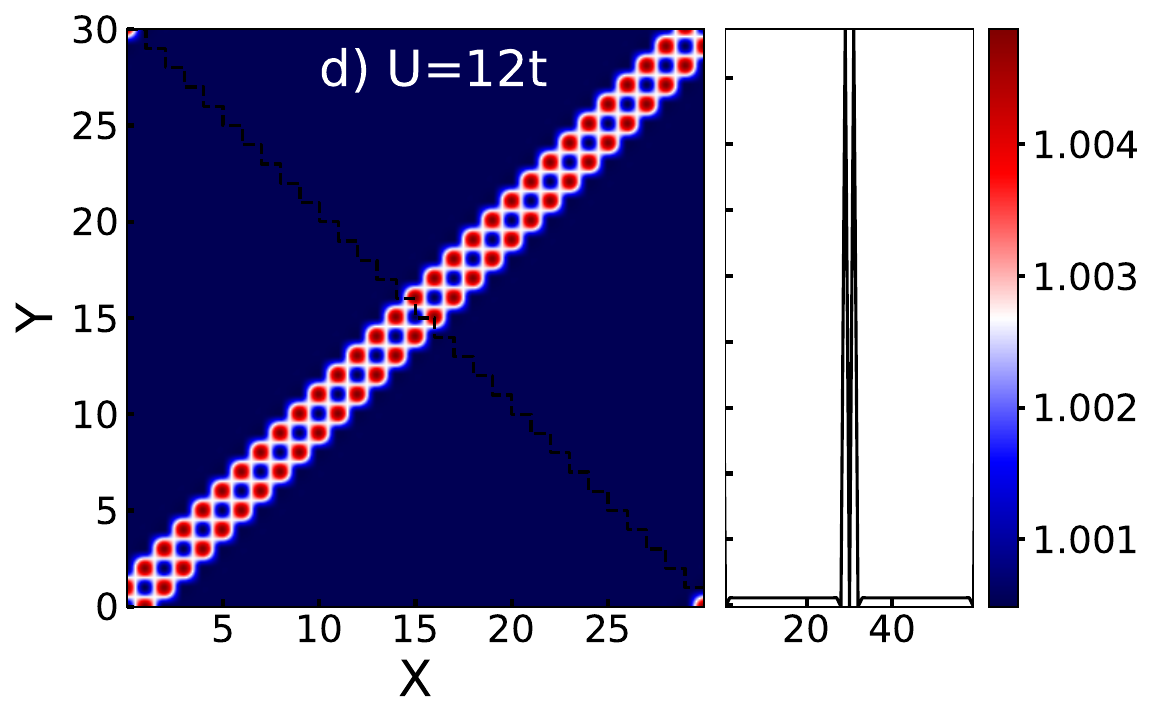}
 \caption{FO in particle denisties ($\bar{n}_{i}$) due to the scattering from  a chain of impurity atoms of equal magnitude 
 $V_{0}=24t$ along the diagonal  direction of the square lattice. Interference effects on FO are shown for a) non-interacting system ($U=0t$), b) $U=2t$, c) $U=5t$, and d) $U=12t$. The model, all other parameters, and the plotting style are the same as in Fig.~\ref{2dtwoimpvert}.
 }
 \label{2ddomainwall}
 \end{figure}

Next, we  study the case where a chain of impurities each of magnitude ($V_{0} = 24t$) is placed along the diagonal direction and  a vertical direction of the square ($31\times31$) lattice. 
This aims to model a domain wall or an interface. 
We investigate the behavior of FO for these two orientations of the chain with the different interactions. 
Firstly, in Fig. \ref{2ddomainwall}  we present the case for the diagonally oriented chain for the non-interacting system and systems with $U=2t,5t$, and $ 12t$. 
The insets shows a projection of the densities along a zigzag line oriented perpendicularly to the diagonal direction. 
FO are visible around the chain and the behavior is similar for the non-interacting system and $U=2t$, like in the other previously presented  cases. 
FO get localized around the chain with the increasing interaction (cf., $U=5t$). 
On the ends of the cut (corners of the lattice) we observe an increase of oscillations, which is  due to the imposed periodic  boundary conditions in a finite (small) system. 
In the case of a Mott insulator with $U=12t$ no FO are visible. 
The chain creates an interface and effectively forms two subsystems separated in space. \\

\begin{figure} [ht!]
 \centering
 \includegraphics[width=0.3\textwidth]{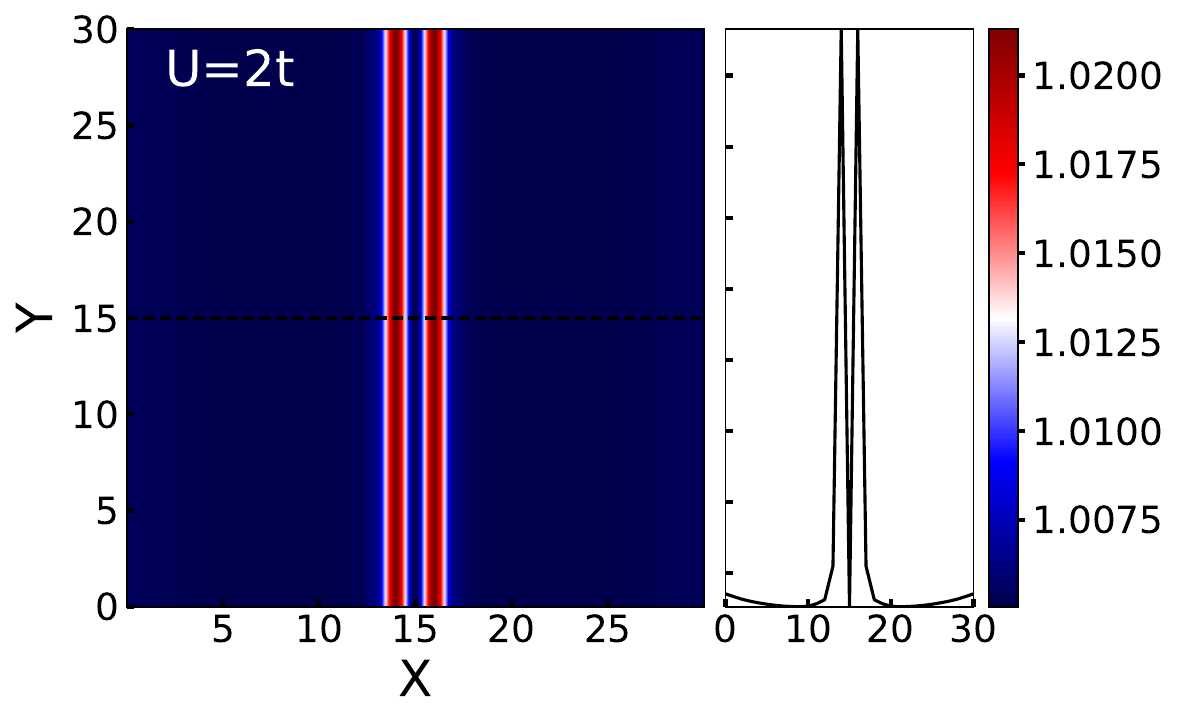} 
 \qquad
 \includegraphics[width=0.3\textwidth]{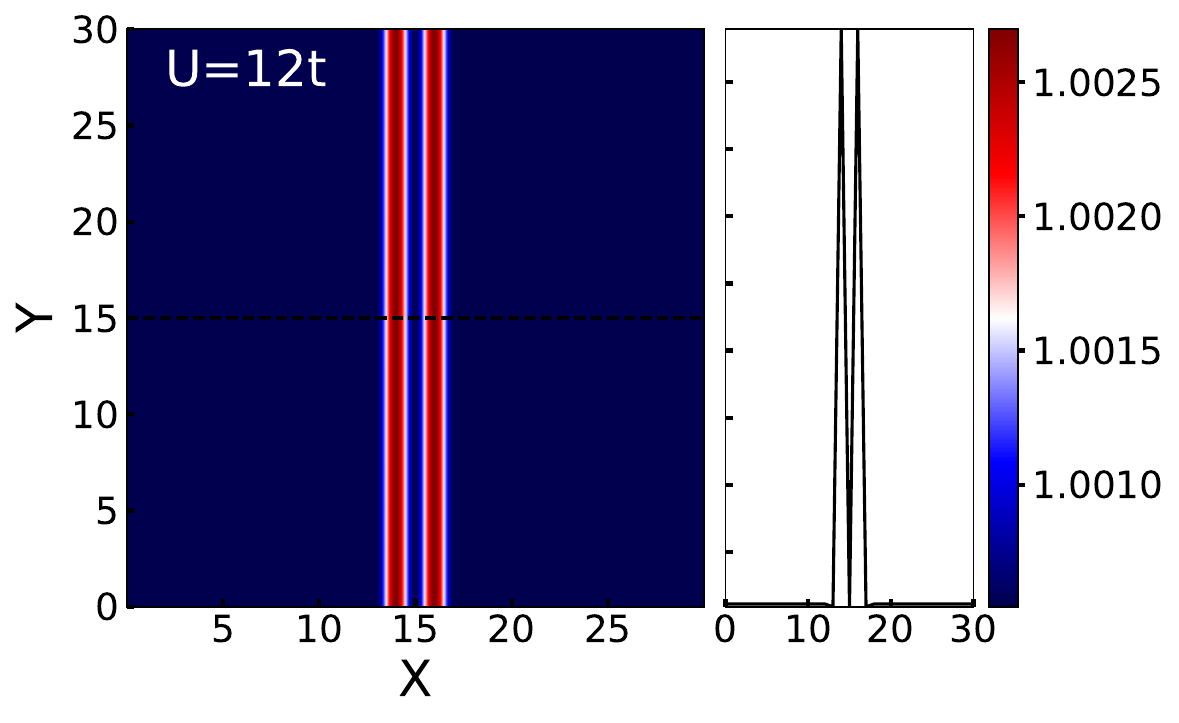}
 \caption{FO in particle denisties ($\bar{n}_{i}$) due to the scattering from  a chain of impurity atoms of equal magnitude 
 $V_{0}=24t$ along the vertical line  of the square lattice. Interference effects on FO are shown for a) $U=2t$, and b)$U=12t$. The model, all other parameters, and the plotting style are the same as in Fig.~\ref{2dtwoimpvert}.
 }
 \label{2ddomainwallvert}
 \end{figure}

 In Fig. \ref{2ddomainwallvert} we show the behavior for the system with the interactions if the chain of impurities is oriented along the vertical direction. 
The inset shows a horizontal cut perpendicular to the chain. 
In contrast, to the previous case we do not see any FO but just a density minimum corresponding to the repulsive potential both for the non-interacting and interacting systems. 
We only show the cases for $U=2t$ and $U=12t$ since the behavior of the system do not change much with the interactions. 
This different behavior as compared to the diagonally oriented chain lies in the geometrical orientation of the impurities with respect to each other.
 If the chain is vertically oriented, each impurity site has two neighboring sites occupied by the impurities. 
 On the other hand, if the chain is diagonally oriented,  each impurity site is completely surrounded by nearest neighboring sites without impurities. 
 Hence, in the latter case the distance between the impurity sites and the sites on a perpendicular cut, measured in Manhattan metric, is always even, in contrast to the former  case, where it is always odd. 
 This difference makes the interference pattern between FO created by each impurity from the chain always constructive in the diagonal case, while it is destructive in the vertical case.\\
 	  
\begin{figure} [ht!]
\centering
\includegraphics[width=0.4\textwidth]{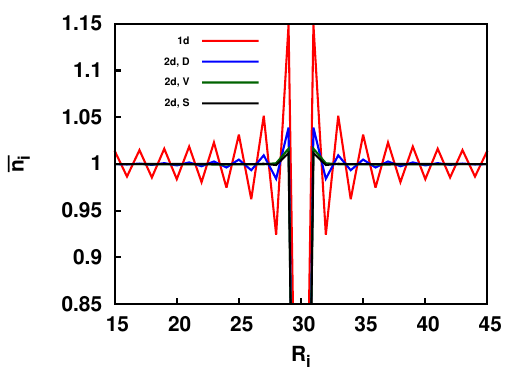} 
\caption{Comparison of FO along:  1d lattice chain with $N_{L}=32$ sites and a single impurity potential $V_{0}=12t$ placed at the center (1d), across the perpendicular cut for a chain of impurities $V_{0}=24t$ along the diagonal chain (D), vertical chain (V), and  a single impurity potential $V_{0}=24t$ placed at the centre of the square lattice (S). We show the non-interacting case $U=0t$.}
\label{1ddomain}
\end{figure}

 In Fig. \ref{1ddomain} we compare the FO  from the diagonally oriented chain  (blue line) and from the vertically oriented chain (green line)  of impurities,  presented above, with the FO of  a one-dimensional lattice having a single impurity potential (red line), with $N_{L}=32$ sites and a  single impurity $V_{0}=12t$ placed in the center. 
We also show the FO  from a  single impurity potential in the square lattice (black line). 
We consider the non-interacting systems for all these cases.
The comparison shows that none of the 2d systems  could be simplified to an assembly of  1d chains with a single impurity potential. 
While the vertically oriented chain shows no oscillations, the decay of FO due to the diagonally oriented chain is not exactly similar to the 1d chain. 
Eventually, the FO from both the vertically and diagonally oriented chains are also quite different compared to a chain from a 2d lattice with a single impurity potential at the center. 
Hence, the substantial role of the geometrical orientation of the chain of impurities on the interference effect   prevents one to simplify this system to an equivalent 1d chains with  single impurity potentials  assembled together. \\

 %%%%%%%%%%%%%%%%%%%%%%%%%%%%%%%%%%%%%%%%%%%%%%%%%%%%%%%%%%%%%%%%%%%%%%%%%%%%%%%%%%%%%%%%%%%%%%%%%%%%%%%%%%%%%%	     

   %%%%%%%%%%%%%%%%%%%%%%%%%%%%%%%%%%%%%%%%%%%%%%%%%%%%%%%%%%%%%%%%%%%%%%%%%%%%%%%%%%%%%%%%%%%%%%%%%%%%%%%%%%%%%%%%
 	
 	\subsection {Extended inhomogeneity}

\begin{figure} [ht!]
\centering
\includegraphics[width=0.3\textwidth]{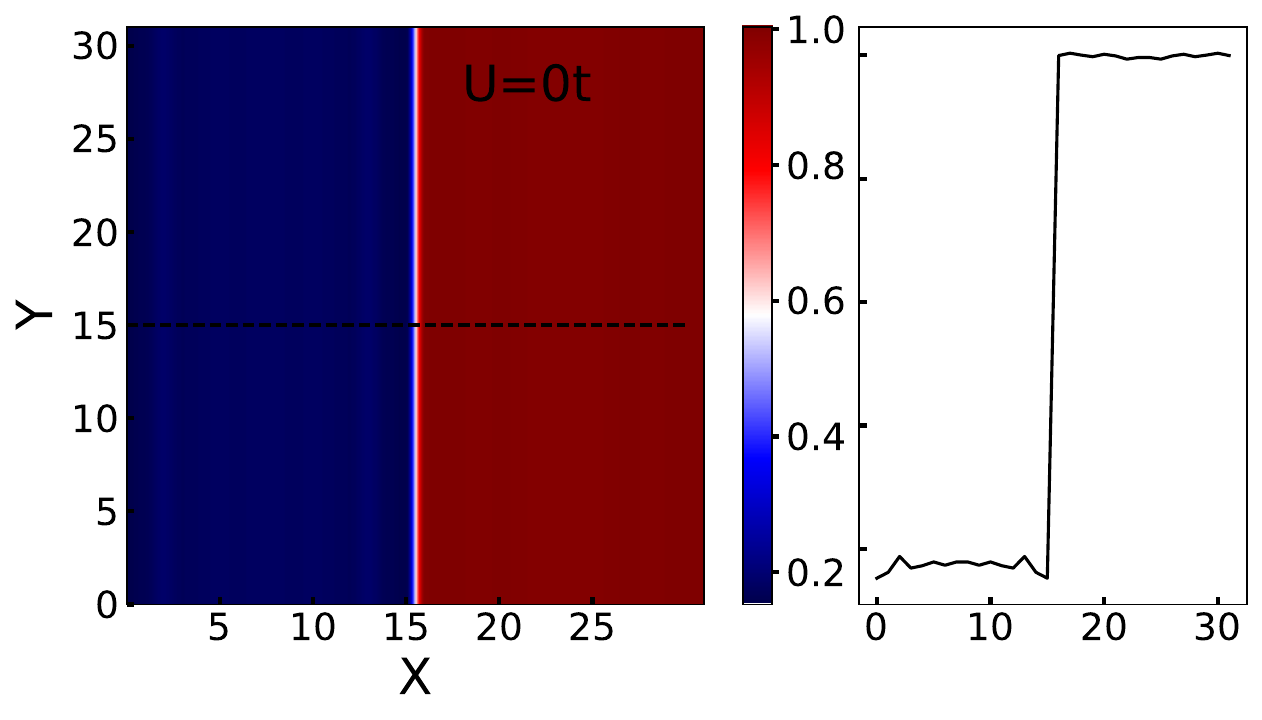} 
\qquad
\includegraphics[width=0.3\textwidth]{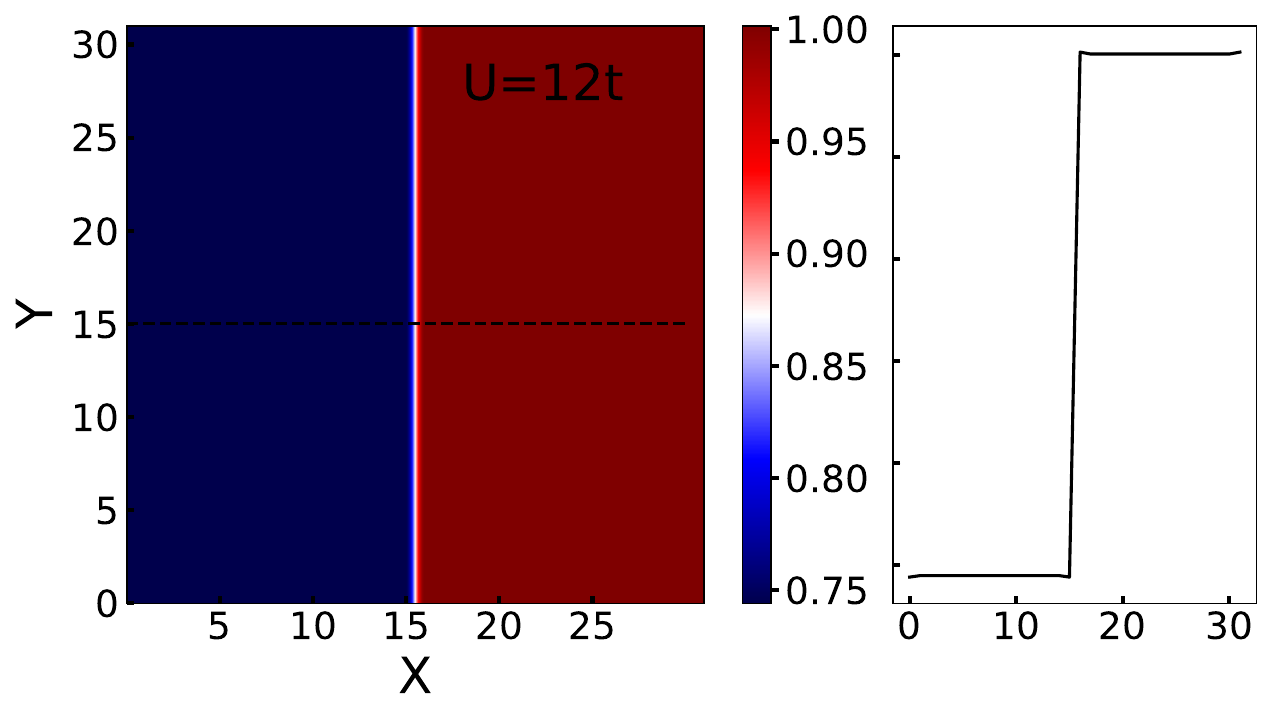}  
\qquad
\includegraphics[width=0.3\textwidth]{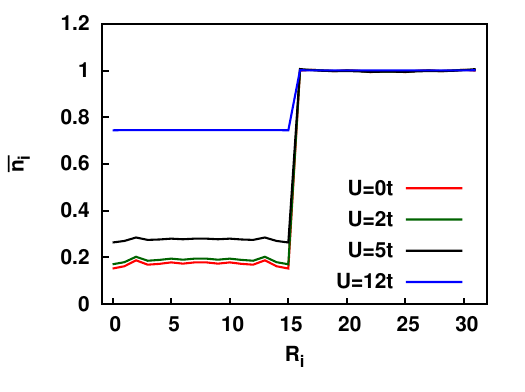}
\caption{Interference effect on FO  due to the scattering from an extended inhomogeneity: 
A step like potential is applied in half of  the square lattice 
($32\times32$), such that on the left side of the vertical symmetry line the potential is $V_{0}=3t$ and on the right side of it $V_{0}=0$. We show the non-interacting case (upper panel) and Mott phase ($U=12t$) (middle panel). The right insets show oscillations on a  cut (dotted line in the main panels)  perpendicular to the step of the potential. In the bottom panel we compare a similar cut for the different interactions $U=0t$, $2t$, $5t$, and  $12t$.}
\label{2dextinhom}
\end{figure}

 	 Next, we apply a step like potential across the square lattice ($32\times32$), i.e. for all the lattice sites with x-coordinates $X_i\leqslant 15a$ the potential is $V_{0}=3t$ and in the rest of the system is $V_{0}=0$. 
	 This potential models an extended inhomogeneity which could correspond to the surface irregularities in materials developed during the process of  cleavage. 
	 In Fig. \ref{2dextinhom} we show local densities in the non-interacting system (upper panel) and in the interacting system with  $U=12t$ (middle panel).
	 The step-like potential divides the lattice into two half-planes with a different average occupation ($\bar{n}_{i}$). 
	 FO is visible for $U=0$, but the period of oscillations differ in the two half-planes as illustrated in the inset, where we show the FO in the  cut perpendicular to the potential edge.
	 Different oscillation periods originate from different uniform densities of particles in each halves of the systems.  
	 Any signature of FO is  absent for the Mott phase ($U=12t$), cf. the bottom panel of Fig. \ref{2dextinhom}. 
	 In Fig. \ref{2dextinhom} (bottom panel) the influence of interactions in this system is studied taking  cuts perpendicular to the step of the potential. 
	 We see that the system becomes more homogeneous and the screening charge decreases with the increasing interaction.  \\
	 
 	 \subsubsection{Going beyond a simple HSEA scheme}

The extended step-like potential discussed here preserves partially a mirror point symmetries of the lattice. 
Therefore, it is natural to extend our earlier HSEA scheme and determine homogeneous self-energies, which correspond to the left side of the system and to the right side independently. 
In the right side the external potential is zero and the corresponding infinite system is at half-filling. 
On the left side a finite potential $V_i=3t$ corresponds to a change of the chemical potential in the corresponding infinite system. 
Therefore, the reference system is away from half filling. \\

We used two different homogeneous self-energies, obtained by solving DMFT equations using the NRG method, for the infinite homogeneous systems and the chemical potentials $\mu=0$ and $\mu=-V_i$, for the left and the right sides respectively. 
Then from solving the Dyson equation (\ref{greenfunction}) we obtain actual densities of particles. \\

\begin{figure} [ht!]
\centering
\includegraphics[width=0.4\textwidth]{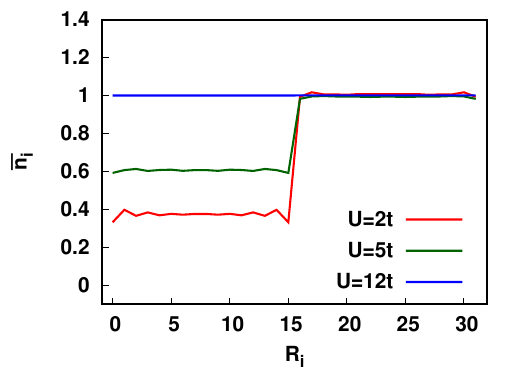} 
\caption{The same cut as in Fig.\ref{2dextinhom} using the self-energy from a  doped Hubbard model.}
\label{doped}
\end{figure}

In the results, shown in Fig.~\ref{doped}, we see that the main difference is in changing the average density on left side of the system. 
However,  shapes and characters of oscillations seems to be very similar to those seen in Fig.~\ref{2dextinhom}. 
Hence we conclude that the HSEA does not take into account very precisely the Hartree (static) term in the self-energy, which on the impurity sites is quite different from the homogeneous case. 
However, regarding how the particle density oscillates around the average density, it seems to be independent of this Hartree term. \\

\begin{figure} [ht!]
\centering
\includegraphics[width=0.4\textwidth]{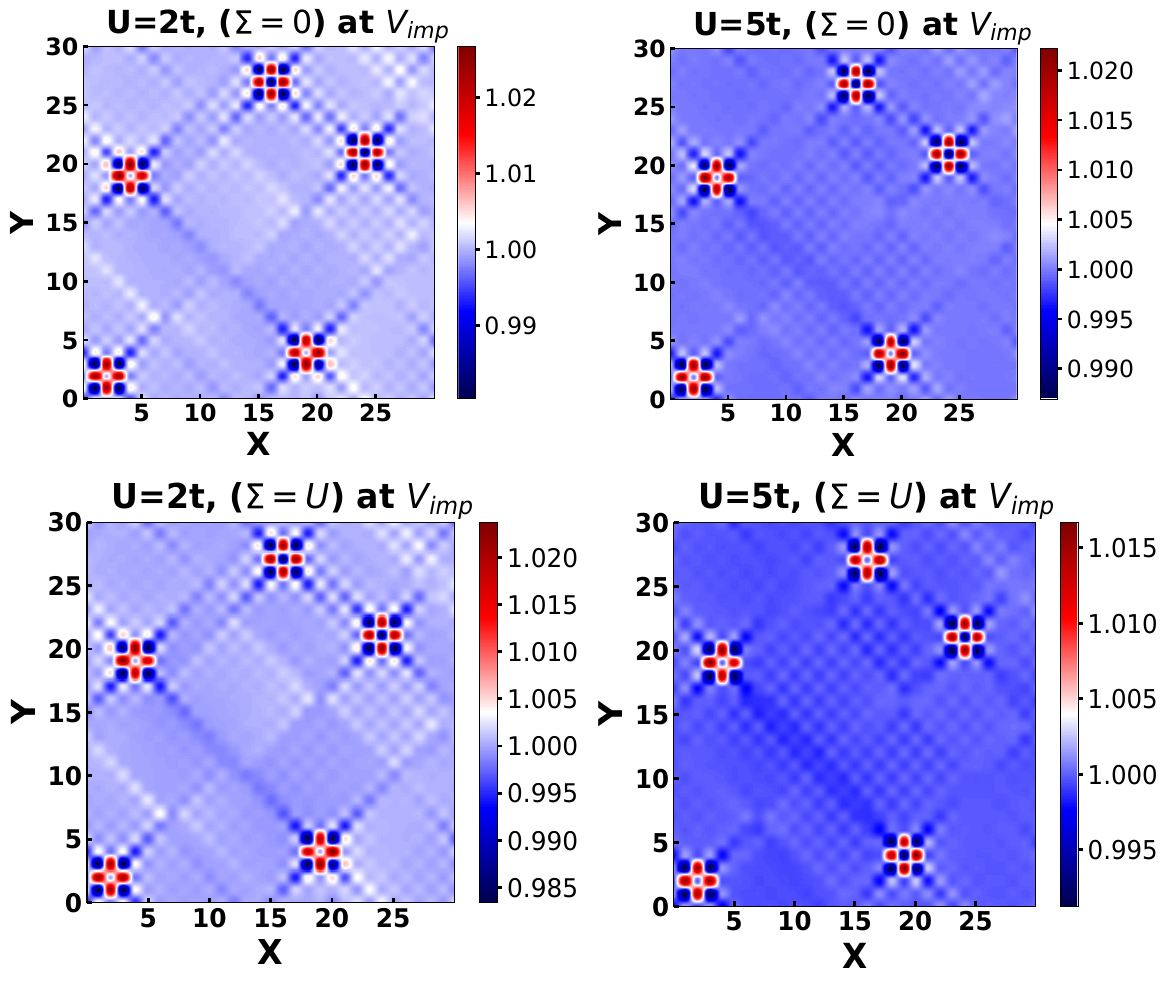} 
\caption{FO in particle densities ($\bar{n}_{i}$)  in the presence of five impurities each of magnitude $V_{1}=V_{2}=V_{3}=V_{4}=V_{5}=10t$ randomly distributed  over the same lattice  sites as in Fig. \ref{2dmultimp}. We put the self-energy to be zero (top panel) and the self-energy to be equal to $U$ (bottom panel) at the impurity sites. 
Otherwise, it is 
We do not see any changes from the HSEA.}
\label{selfV}
\end{figure}

To make sure that our reasoning is correct we determined the particles densities in the case of few impurities randomly distributed on the lattice, as in Fig.~\ref{2dmultimp} in Sec.~IIIB, but with a modified HSEA scheme. I.e.,  on the impurity sites either the self-energy is taken to be zero, cf. upper panels in Fig.~\ref{selfV}, or the self-energy is equal to the interaction, cf. lower panels in  Fig.~\ref{selfV}. 
Comparing Figs.~\ref{selfV}~and~\ref{2dmultimp} we conclude that the FO oscillations are still the same. \\

%Hence, we need to do a full R-DMFT to see how the results change from HSEA. Our insights on the expected changes is presented in Section IV.
 
\section{Conclusions}
 \label{summary}
 
 We have studied the interference effects in FO due to the scattering from two impurities, multiple impurities, extended inhomogeneities in  non-interacting and interacting fermion systems. 
 On comparing the FO in the presence of a single impurity we see that in two impurity systems, the additional impurity  induces interference effects on FO. The interference maxima and minima change with the relative position of the impurities up to a certain cross-over distance beyond which the impurities behave independently. 
 At half filling, the interaction does not change the position of maxima and minima but reduces their intensity and consequently the interference effects. The screening charge shows an oscillatory behavior with the even and odd lattice spacing between the impurities along a vertical column. A more complex pattern is seen in the presence of multiple impurities but the FO still localize around the impurities with the increase in interaction. In case of extended inhomogeneities the system also becomes more homogeneous with increasing the interaction. In case of a chain of impurities in the square lattice FO is present for a diagonal chain while absent for a vertical chain due to constructive and destructive of FO in these two geometries respectively. In all the models of the impurity potential no FO or interference effects are seen in the Mott insulating phase.\\

% \subsubsection*{Expected changes beyond HSEA using the full R-DMFT}
 %   \bcnote{This section is newly written.}
   We have used a homogeneous self-energy approximation based on DMFT for most of our studies where the inhomogeneous part of the self-energy due to the contribution from the impurities is neglected.
   However, in the case of the extended inhomogeneity we show that the inclusion of the inhomogeneous part of the self-energy do not qualitatively alter the pattern of FO. It is a promising future work to comapare the results of HSEA and full R-DMFT accounting for the inhomogeneous part of the self-energy for the models of multiple discreet impurity potential. We do not expect significant changes with the single site R-DMFT since a possible change in the density of particles due to a change of the self-energy by the inhomogeneous potential is a higher order effect beyond the linear response regime. Hence, one can further probe beyond the single site DMFT taking the spatial correlations, non-local part of the self-energy into account and see if the positions of interference maxima/minima are altered by the interactions. Our model studies should further motivate a realistic modeling of FO in real materials like transition metal oxides, transition metal dichalcogenides using LDA+DMFT in real space and its validation through STM experiments.\\
   
   \begin{acknowledgments}
%    This work was supported by Foundation for Polish Science (FNP) through the TEAM/2010-6/2 project, co-financed by the EU European Regional Development Fund. 
    
    We thank D. Vollhardt for motivating discussions. Also a comment from an anonymous referee concerning HSEA is acknowledged. 
    BC thanks  J. Koloren{\v{c}}, V. Pokorn{\'y}, and J. Mravlje for discussions. 
    Computing facilities provided by the Czech National Grid Infrastructure MetaCentrum (project CESNET LM2015042) is gratefully acknowledged. BC was partially supported by the Slovenian Research Agency ARRS by Project No. J1-2458 and P1-0044.
    
    \end{acknowledgments}
    
%     \appendix


\begin{thebibliography}{}
\bibitem{Friedel52} J.~Friedel, Phil. Mag. {\bf 43}, 153 (1952).

\bibitem{Friedel58} J.~Friedel, Nuovo Cimento Supp. {\bf 7}, 287 (1958).

\bibitem{alloul_friedel_2012} H. Alloul, 
J. Supercond. Nov. Magn. {\bf 25}, 585 (2012).

\bibitem{1742-6596-592-1-012059} B. Chatterjee and K. Byczuk, 
 J. Phys.: Conf. Ser. {\bf 592}, 012059 (2015).
 
 \bibitem{byczuk2019t}K. Byczuk, B. Chatterjee, D. Vollhardt,
Euro. Phys J B {\bf 92}, 23 (2019).

\bibitem{chatterjee2019real}B. Chatterjee, J. Skolimowski, K. Makuch, K. Byczuk,
Phys. Rev. B {\bf 100}, 115118 (2019).

\bibitem{Crommie93} M. F. Crommie, C. P. Lutz, and D. M. Eigler, Science {\bf 262}, 218 (1993).

\bibitem{Eigler90} D. M. Eigler and E. K. Schweizer, Nature (London) {\bf 344}, 524 (1990).

\bibitem{Kanisawa01} K. Kanisawa, M. Butcher, H. Yamaguchi, and Y. Hirayama,  Phys. Rev. Lett {\bf 86}, 3384 (2001).

\bibitem{Binnig82} G. Binnig, H. Rohrer,  Ch. Gerber, and E. Weibl, Appl. Phys. Lett. {\bf 40}, 178 (1982);  Phys. Rev. Lett. {\bf 49}, 57 (1982); {\it ibid}. {\bf 50}, 120 (1983). 
\bibitem{clark2014energy} K. W. Clark, X-G. Zhang, G. Gu, J. Park, G. He, R. M. Feenstra, and A.-P. Li, 
Phys. Rev. X {\bf 4}, 011021 (2014).

\bibitem{chico} Chico L et. al, Acta Phys. Pol. A, {\bf 114} (2008).

\bibitem{kolesnychenko2005surface} O. Y. Kolesnychenko, G. M. M. Heijnen, A. K. Zhuravlev, R. de Kort, M. I. Katsnelson, A. I. Lichtenstein, H. van Kempen, Phys. Rev. B {\bf 72}, 085456 (2005).

 \bibitem{egger1995friedel} R. Egger, and H. Grabert, Phys. Rev. Lett {\bf 75}, 3505 (1995).
 
   \bibitem{schuster2004local}C. Schuster, and P. Brune, Phys. Status Solidi B{\bf 241}, 2043 (2004).
   
 \bibitem{vieira2008friedel} D.Vieira, H. J. P. Freire, V. L. Campo Jr, and K. Capelle, J. Magn. Magn. {\bf 320}, 418 (2008).
 
  \bibitem{simion2005friedel} G. E. Simion, and G. F. Giuliani, Phys. Rev. B {\bf 72}, 045127 (2005).
  
  \bibitem{ziegler1998friedel} W. Ziegler, H. Endres, and W. Hanke, 
 Phys. Rev. B {\bf 58}, 4362 (1998).
 
 \bibitem{affleck_friedel_2008} I. Affleck, L. Borda, and H. Saleur, 
Phys. Rev. B {\bf 77}, 180404 (2008).
 
 \bibitem{vollhardt2012Dynamical}  D. Vollhardt, K. Byczuk, and M. Kollar, 
in {\em Strongly Correlated Systems},  ed. by A. Avella and F. Mancini,  p.~ 203 (Springer, 2012).

%\bibitem{vollhardt2010dynamical} D. Vollhardt, A. Avella, and F. Mancini, AIP Conference Proceedings {\bf 1297}, 339 (2010).

\bibitem{vollhardt_investigation_1993} D. Vollhardt, in
{\it Correlated Electron Systems}, Ed. V.J. Emery, p.~57 (World Scientific, Singapore, 1993). 

\bibitem{kotliar2004strongly} G. Kotliar, and D. Vollhardt,  
Phys. Today {\bf 57}, 53 (2004).

\bibitem{byczuk2008dynamical} K. Byczuk, in
{\em Condensed Matter Physics In The Prime Of The 21st Century: Phenomena, Materials, Ideas, Methods},  ed. by Janusz Jedrzejewski, p.~1 (World Scientific, 2008).

\bibitem{titvinidze_dynamical_2012}I. Titvinidze, A. Schwabe, N. Rother, and M. Potthoff, 
Phys. Rev. B {\bf 86}, 075141 (2012).

\bibitem{0953-8984-9-35-010} V. I. Anisimov, A. I. Poteryaev, M. A. Korotin, A. O. Anokhin, and G. Kotliar, 
J. Phys. Condens {\bf 9}, 7359 (1997).

\bibitem{Helmes08} R.~W.~Helmes, T.~Costi, and A.~Rosch, Phys. Rev. Lett. {\bf 100}, 056403 (2008); {\em ibid}. {\bf 101}, 066802 (2008). 

\bibitem{Snook08} M.~Snoek, I.~Titvinidze, C.~Toke, K.~Byczuk, and W.~Hofstetter, New J. Phys. {\bf 10} 093008 (2008). 

\bibitem{potthoff1999metallic} M. Potthoff, and W. Nolting, Phys. Rev. B {\bf 60}, 7834 (1999). 


\bibitem{suarez2020two} M. Y. Su{\'a}rez-Villagr{\'a}n, N. Mitsakos, T. H. Lee, V. Dobrosavljevi{\'c}, J. H. Miller,  E. Miranda,  Phys. Rev. B {\bf 101}, 235112 (2020).



\bibitem{freericks2006transport} J. K. Freericks, in 
{\em Transport in Multilayered Nanostructures: The Dynamical Mean-field Theory Approach},  (Imperial College Press, 2006).

\bibitem{grosu2008friedel} I. Grosu, and L. Tugulan, J. Supercond. Nov. Magn {\bf 21}, 65 (2008).

\bibitem{liu2003two} Y. L. Liu, Phys. Rev. B {\bf 68}, 155116 (2003).

\bibitem{derry2015quasiparticle} P. G. Derry, A. K. Mitchell, D. E. Logan, 
 Phys. Rev. B {\bf 92}, 035126 (2015).
 \bibitem{mitchell2015multiple} A. K. Mitchell, P. G. Derry, Andrew K D. E. Logan, 
 Phys. Rev. B {\bf 91}, 235127 (2015).
 
 \bibitem{schuler2016many} M. Sch{\"u}ler, S. Barthel, M. Karolak, A. I. Poteryaev, A. I. Lichtenstein, M. I. Katsnelson, G. Sangiovanni, and T. O. Wehling, 
 Phys. Rev. B {\bf 93}, 195115 (2016).
 
  \bibitem{kolorenvc2015electronic} J. Koloren{\v{c}}, A. B. Shick, and A. I. Lichtenstein Phys. Rev. B {\bf 92}, 085125 (2015).
  
\bibitem{choi2017mapping} D. J. Choi, C. R. Verd{\'u}, J. de Bruijckere, M. M. Ugeda, N. Lorente, and J. I. Pascual, Nat. Comm. B {\bf 8}, 1 (2017).

\bibitem{hubbard_electron_1963} J. Hubbard, 
Proc. Royal Society of London A: Mathematical, Physical and Engineering Sciences {\bf 276}, 238 (1963). 
 \bibitem{gutzwiller1963effect} M. C. Gutzwiller, 
 Phys. Rev. Lett. {\bf 10}, 159 (1963).
 
 \bibitem{kanamori1963electron} J. Kanamori, 
  Prog. Theor. Phys. {\bf 30}, 275 (1963).
 
  \bibitem{rickayzen1980green} G. Rickayzen, 
{\em Green's functions and condensed matter},  (London; New York Academic Press, 1980).

\bibitem{bulla_zero_1999} R. Bulla, 
Phys. Rev. Lett. {\bf 83}, 136 (1999).

\bibitem{vzitkonrg} R. {\v{Z}}itko, 
 www.nrgljubljana.ijs.si (2014).
 
 \bibitem{costi1990new} T. A. Costi, and A. Hewson, 
Physica B: Cond Matt {\bf 163}, 179 (1990).

\end{thebibliography}
\end{document}